\documentclass[twocolumn]{aastex631}
\usepackage{CJK}
\usepackage{xcolor}

\newcommand{\kms}{km s$^{-1}$}

\begin{document}
\begin{CJK}{UTF8}{mj}
\title{The FU\,Ori outburst of PR Ori B}

\author[0000-0003-1894-1880]{Carlos Contreras Pe\~{n}a}
\affiliation{Department of Physics and Astronomy, Seoul National University, 1 Gwanak-ro, Gwanak-gu, Seoul 08826, Korea}
\affiliation{Research Institute of Basic Sciences, Seoul National University, Seoul 08826, Republic of Korea}
\author[0000-0002-3632-1194]{Zs\'ofia Nagy}
\affiliation{Konkoly Observatory, HUN-REN Research Centre for Astronomy and Earth Sciences, MTA Centre of Excellence, Konkoly-Thege Mikl\'os \'ut 15-17, 1121 Budapest, Hungary}
\author[0000-0002-7154-6065]{Gregory Herczeg (沈雷歌）}
\affiliation{Kavli Institute for Astronomy and Astrophysics, Peking University, Yiheyuan Lu 5, Haidian Qu, 100871 Beijing, Peoples Republic of China}
\affiliation{Department of Astronomy, Peking University, Yiheyuan 5, Haidian Qu, 100871 Beijing, China}
\author[0000-0003-3119-2087]{Jeong-Eun Lee (이정은)}
\affiliation{Department of Physics and Astronomy, Seoul National University, 1 Gwanak-ro, Gwanak-gu, Seoul 08826, Korea}
\affiliation{SNU Astronomy Research Center, Seoul National University, 1 Gwanak-ro, Gwanak-gu, Seoul 08826, Korea}
\author[0000-0001-7157-6275]{\'Agnes K\'osp\'al}
\affiliation{Konkoly Observatory, HUN-REN Research Centre for Astronomy and Earth Sciences, MTA Centre of Excellence, Konkoly-Thege Mikl\'os \'ut 15-17, 1121 Budapest, Hungary}

\author[0000-0001-6015-646X]{P\'eter \'Abrah\'am}
\affiliation{Konkoly Observatory, HUN-REN Research Centre for Astronomy and Earth Sciences, MTA Centre of Excellence, Konkoly-Thege Mikl\'os \'ut 15-17, 1121 Budapest, Hungary}
\affiliation{Department of Astrophysics, University of Vienna, T\"urkenschanzstrasse 17, 1180 Vienna, Austria}
\author[0000-0002-6773-459X]{Doug Johnstone}
\affiliation{NRC Herzberg Astronomy and Astrophysics, 5071 West Saanich Rd, Victoria, BC, V9E 2E7, Canada}
\affiliation{Department of Physics and Astronomy, University of Victoria, Victoria, BC, V8P 5C2, Canada}
\author[0000-0001-8174-1932]{Bo Reipurth}
\affiliation{Institute for Astronomy, University of Hawaii at Manoa, 640 N. Aohoku Place, HI 96720, USA}
\affiliation{Planetary Science Institute, 1700 E Fort Lowell Rd, Suite 106, Tucson, AZ 85719, USA}
\author[0000-0002-2523-3762]{Chul-Hwan Kim (김철환)}
\affiliation{Department of Physics and Astronomy, Seoul National University, 1 Gwanak-ro, Gwanak-gu, Seoul 08826, Korea}
\author[0000-0001-9515-3584]{Hyun-Il Sung (성현일)}
\affiliation{Korea Astronomy and Space Science Institute, 61-1 Whaam-dong, Yuseong-gu, Daejeon 305-348, Korea}
\author[0009-0006-7226-711X]{Yihan Li}
\affiliation{Department of Astronomy, Peking University, Yiheyuan 5, Haidian Qu, 100871 Beijing, China}
\author{Jiaze Wang}
\affiliation{Department of Astronomy, Peking University, Yiheyuan 5, Haidian Qu, 100871 Beijing, China}
\author{Rui Wu}
\affiliation{Department of Astronomy, Peking University, Yiheyuan 5, Haidian Qu, 100871 Beijing, China}
\author[0000-0002-6394-8013]{Foteini Lykou}
\affiliation{Konkoly Observatory, HUN-REN Research Centre for Astronomy and Earth Sciences, MTA Centre of Excellence, Konkoly-Thege Mikl\'os \'ut 15-17, 1121 Budapest, Hungary}
\author[0009-0000-4741-7055]{Mizna Ashraf}
\affiliation{Department of Physics, Indian Institute of Science Education and Research Tirupati, Yerpedu, Tirupati - 517619, Andhra Pradesh, India}
\author[0000-0001-8135-6612]{John Bally}
\affiliation{Center for Astrophysics and Space Astronomy, Department of Astrophysical and Planetary Sciences University of Colorado, Boulder, CO 80389, USA;}
\author[0000-0002-4283-2185]{Fernando Cruz-Sa\'enz de Miera}
\affiliation{Institut de Recherche en Astrophysique et Planétologie, Université de Toulouse, UT3-PS, OMP, CNRS, 9 av. du Colonel Roche, 31028 Toulouse Cedex 4, France}
\affiliation{Konkoly Observatory, HUN-REN Research Centre for Astronomy and Earth Sciences, MTA Centre of Excellence, Konkoly-Thege Mikl\'os \'ut 15-17, 1121 Budapest, Hungary}
\author[0000-0001-8060-1321]{Min Fang}
\affiliation{Purple Mountain Observatory, Chinese Academy of Sciences, Nanjing 210023, China}
\affiliation{School of Astronomy and Space Sciences, University of Science and Technology of China, Hefei 230026, China}
\author[0000-0002-5261-6216]{Eleonora Fiorellino}
\affiliation{Alma Mater Studiorum, Universit\'a di Bologna, Dipartimento di Fisica e Astronomia ``AugustoRighi,'' ViaGobetti 93/2, I-40129 Bologna, Italy}
\affiliation{INAF Osservatorio Astronomico di Trieste, via Tiepolo11, I-34143 Trieste, Italy}
\author[0000-0002-4223-103X]{Christoffer Fremling}
\affiliation{Caltech Optical Observatories, California Institute of Technology, Pasadena, CA 91125, USA}
\affiliation{Division of Physics, Mathematics and Astronomy, California Institute of Technology, Pasadena, CA 91125, USA}
\author[0000-0002-7035-8513]{Teresa Giannini}
\affiliation{INAF-Osservatorio Astronomico di Roma, via di Frascati 33, 00078, Monte Porzio Catone, Italy}
\author[0000-0003-4908-4404]{Jessy Jose}
\affiliation{Department of Physics, Indian Institute of Science Education and Research Tirupati, Yerpedu, Tirupati - 517619, Andhra Pradesh, India}
\author[0000-0002-5619-4938]{Mansi Kasliwal}
\affiliation{Division of Physics, Mathematics and Astronomy, California Institute of Technology, Pasadena, CA 91125, USA}
\author[0000-0002-7538-5166]{M\'aria Kun}
\affiliation{Konkoly Observatory, HUN-REN Research Centre for Astronomy and Earth Sciences, MTA Centre of Excellence, Konkoly-Thege Mikl\'os \'ut 15-17, 1121 Budapest, Hungary}
\author{Ho-Gyu Lee (이호규)}
\affiliation{Korea Astronomy and Space Science Institute, 776 Daedeok-daero, Yuseong, Daejeon 34055, Korea}

\author[0000-0001-5018-3560]{Michal Siwak}
\affiliation{Mt. Suhora Astronomical Observatory, University of the National Education Commission, ul. Podchora\.zych 2, 30-084 Krak{\'o}w, Poland}
\affiliation{Konkoly Observatory, HUN-REN Research Centre for Astronomy and Earth Sciences, MTA Centre of Excellence, Konkoly-Thege Mikl\'os \'ut 15-17, 1121 Budapest, Hungary}
\author[0000-0002-3648-433X]{M\'at\'e Szil\'agyi}
\affiliation{Konkoly Observatory, HUN-REN Research Centre for Astronomy and Earth Sciences, MTA Centre of Excellence, Konkoly-Thege Mikl\'os \'ut 15-17, 1121 Budapest, Hungary}
\author[0000-0001-6216-0462]{Sung-Yong Yoon (윤성용)}
\affiliation{Korea Astronomy and Space Science Institute, 776 Daedeok-daero, Yuseong, Daejeon 34055, Korea}
\affiliation{School of Space Research, Kyung Hee University, 1732, Deogyeong-daero, Giheung-gu, Yongin-si, Gyeonggi-do 17104, Republic of Korea}
\author{Zs\'ofia Bora}
\affiliation{Konkoly Observatory, HUN-REN Research Centre for Astronomy and Earth Sciences, MTA Centre of Excellence, Konkoly-Thege Mikl\'os \'ut 15-17, 1121 Budapest, Hungary}
\author{Borb\'ala Cseh}
\affiliation{Konkoly Observatory, HUN-REN Research Centre for Astronomy and Earth Sciences, MTA Centre of Excellence, Konkoly-Thege Mikl\'os \'ut 15-17, 1121 Budapest, Hungary}
\affiliation{MTA-ELTE Lend{\"u}let "Momentum" Milky Way Research Group, Hungary}
\author{\'Agoston Horti-D\'avid}
\author[0009-0002-0157-4228]{Zs\'oka Horv\'ath}
\affiliation{Konkoly Observatory, HUN-REN Research Centre for Astronomy and Earth Sciences, MTA Centre of Excellence, Konkoly-Thege Mikl\'os \'ut 15-17, 1121 Budapest, Hungary}
\affiliation{ELTE E\"otv\"os Lor\'and University, Institute of Physics and Astronomy, P\'azm\'any P\'eter s\'et\'any 1/A, Budapest, Hungary}
\author[0000-0001-5203-434X]{Andr\'as Peter Jo\'o}
\affiliation{Konkoly Observatory, HUN-REN Research Centre for Astronomy and Earth Sciences, MTA Centre of Excellence, Konkoly-Thege Mikl\'os \'ut 15-17, 1121 Budapest, Hungary}
\author[0000-0002-1663-0707]{Csilla Kalup}
\author[0009-0007-3760-515X]{Kl\'ara Lelkes}
\affiliation{Konkoly Observatory, HUN-REN Research Centre for Astronomy and Earth Sciences, MTA Centre of Excellence, Konkoly-Thege Mikl\'os \'ut 15-17, 1121 Budapest, Hungary}
\affiliation{ELTE E\"otv\"os Lor\'and University, Institute of Physics and Astronomy, P\'azm\'any P\'eter s\'et\'any 1/A, Budapest, Hungary}
\author{Andr\'as P\'al}
\author{Zsolt Reg\'aly}
\author[0000-0003-0926-3950]{Kriszti\'an S\'arneczky}
\author[0000-0002-3658-2175]{B\'alint Seli}
\author[0000-0001-7806-2883]{\'Ad\'am S\'odor}
\affiliation{Konkoly Observatory, HUN-REN Research Centre for Astronomy and Earth Sciences, MTA Centre of Excellence, Konkoly-Thege Mikl\'os \'ut 15-17, 1121 Budapest, Hungary}
\author[0009-0007-1015-0327]{Norton O. Szab\'o}
\affiliation{Konkoly Observatory, HUN-REN Research Centre for Astronomy and Earth Sciences, MTA Centre of Excellence, Konkoly-Thege Mikl\'os \'ut 15-17, 1121 Budapest, Hungary}
\affiliation{ELTE E\"otv\"os Lor\'and University, Institute of Physics and Astronomy, P\'azm\'any P\'eter s\'et\'any 1/A, Budapest, Hungary}
\author{Zs\'ofia Szab\'o}
\author[0000-0002-1698-605X]{R\'obert Szak\'ats}
\author{N\'ora Tak\'acs}
\affiliation{Konkoly Observatory, HUN-REN Research Centre for Astronomy and Earth Sciences, MTA Centre of Excellence, Konkoly-Thege Mikl\'os \'ut 15-17, 1121 Budapest, Hungary}

\author[0000-0002-4147-3846]{Miguel Vioque}
\affiliation{European Southern Observatory, Karl-Schwarzschild-Str. 2, 85748 Garching bei M\"{u}nchen, Germany}

\correspondingauthor{Carlos Contreras Pena, cecontrep@gmail.com and Gregory Herczeg, gherczeg1@gmail.com}

\begin{abstract}
We report the discovery of a nearby FU\,Ori-type outburst (FUor), PR Ori B, in the L1641 cluster of the Orion star-forming region. The high-amplitude variability was first identified in the NEOWISE (3-5 $\mu$m) photometry of the unresolved PR Ori binary system. Long-term, resolved optical photometric monitoring demonstrates that PR Ori B is the driver of a $\Delta G=5~$mag outburst, while PR Ori A has remained constant over the last 20 years. The near-IR spectrum of PR Ori B changes from a late K-type spectral type during quiescence to a viscously heated disk during outburst, including deep absorption in $^{12}$CO and H$_2$O bands.  The optical spectrum also exhibits features that are commonly associated with FUors, including P Cygni profiles in \ion{Na}{1} D lines and absorption in the \ion{Ca}{2} infrared triplet.  
The luminosity of the outburst (L$_{acc}\sim$30--40 L$_\odot$) is similar to that commonly observed in FUors. The comparison of Spitzer/IRS and VLT/VISIR spectroscopy shows some evidence of silicate crystallisation during the outburst. PR Ori B is one of the closest and brightest FUors discovered over the last few years, only one magnitude fainter than the archetype of the class FU\,Ori. The proximity and brightness will allow for future high angular resolution observations to probe the physics of the inner disk and to evaluate changes in the disk due to the increased luminosity. 
\end{abstract}


\section{Introduction} \label{sec:intro}
FU\,Ori eruptions, also known as FUors, are accretion bursts up to $\sim10^{-4} $~M$_{\odot}$~yr$^{-1}$ that can last for over one hundred years \citep[see review by][]{1996Hartmann}. These outbursts are thought to play a critical role in the growth of stars and the chemistry of protoplanetary disks \citep[see review by][]{2023Fischer}. Historically FU\,Ori outbursts were very rare, with only about 10 known as of 2010 \citep[e.g.][]{2014Audard}.  The past decade of monitoring surveys \citep[PTF/ZTF, {\it Gaia}, ASAS-SN, VVV/VVVX, and NEOWISE, ][]{2011Covey,2017Contreras_a,2018Hillenbrand,2020Szegedi-Elek,2023Nagy,2023Contreras_b,2024Guo,2025Siwak} has revealed roughly 60 more FU\,Ori or FU\,Ori-like bursts \citep{2025Contreras_b}. Most of these eruptions, however, are located at distances further than 1 kpc, too distant to study at high spatial and spectral resolution, and lack pre-burst spectroscopy of the star and disk.

\begin{figure}
	\resizebox{0.47\textwidth}{!}{\includegraphics[angle=0]{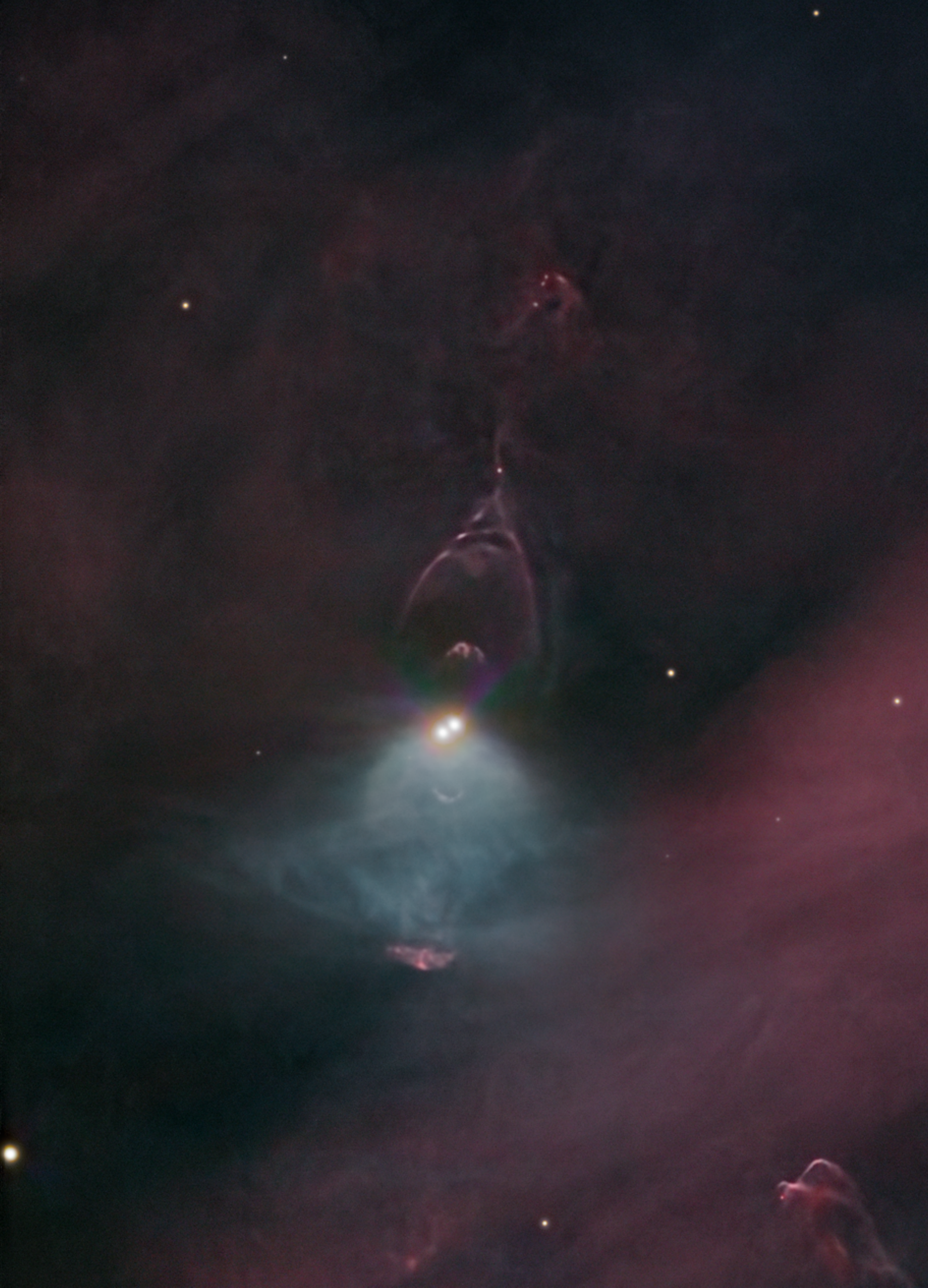}}
	 \caption{Optical image of the PR Ori binary and the outflow lobe, courtesy of Mike Selby (\url{https://throughlightandtime.com/lrgb-nebulaother/}).}
     
	 \label{fig:image}
\end{figure}

In this paper, we analyze the FU\,Ori eruption associated with PR Ori B, discovered in spectroscopic follow-up of NEOWISE variables \citep{2021Park} and initially announced by \citet{2024contrerasATEL}.  PR Ori is a multiple system (see Fig.~\ref{fig:image}) located in the L1641-N cluster in Orion at a {\it Gaia} DR3\footnote{We adopt the cluster distance because PR Ori A and PR Ori B have {\it Gaia} astrometry that are consistent with cluster membership but with a high RUWE, indicating that the parallax is unreliable.} distance of 388.4 pc (see analysis of Gaia DR3 parallaxes by \citealt{2024Froebrich}).
Because FU\,Ori outbursts are rare and cannot be predicted, little information is known about progenitors \citep{herbig1958,1977Herbig,herczeg2025}.  One of the powerful aspects of the PR Ori B outburst is the extensive set of pre-outburst observations, most presented by \citet{2018Reipurth}.

In \S \ref{sec:obs} we describe the photometry and spectroscopy used in this paper. In \S \ref{sec:hist}, we provide a historical overview of the PR Ori system. In \S \ref{sec:pr_erup}-\ref{sec:irspec} we analyze the photometry to establish evidence for the eruption and the spectroscopic interpretation that the eruption is an FU\,Ori outburst. In \S \ref{sec:pr_properties} we then describe the properties of the FU\,Ori burst and progenitor, and in \S \ref{sec:pr_importance} we summarize the importance of this outburst.

\begin{figure*}
	\resizebox{\textwidth}{!}{\includegraphics[angle=0]{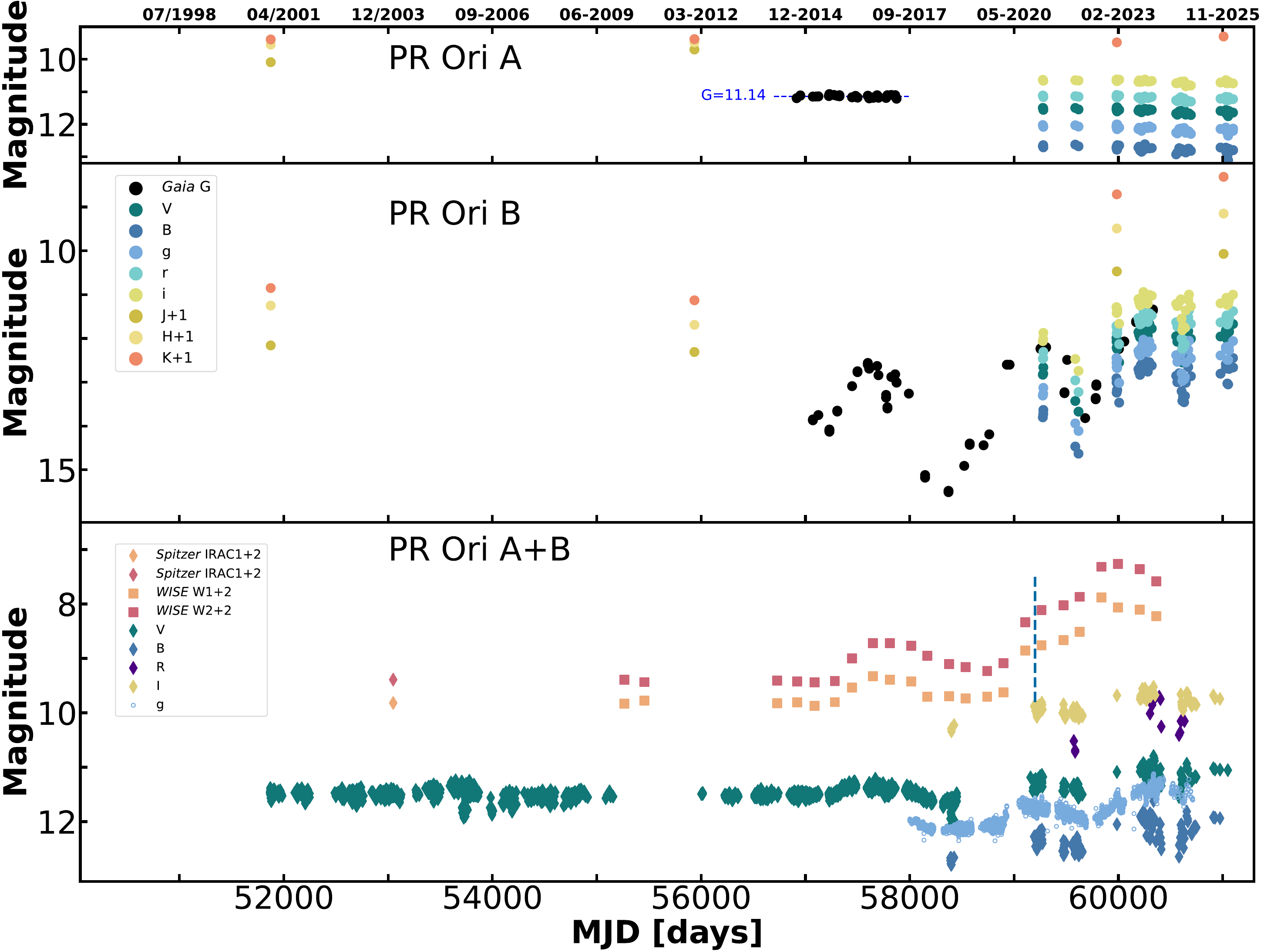}}
	 \caption{(Top) Photometry of PR Ori A (from {\it Gaia} DR3 and the RC80 telescope). (Middle) Photometry of PR Ori B (from the {\it Gaia} alerts and the RC80 telescope). (Bottom) Optical ($B$, $V$, $R$, $I$ and $g$) and mid-IR  ({\it Spitzer} IRAC1 and IRAC2, and {\it WISE} W1 and W2) light curve of the PR Ori system (unresolved photometry). The classification as a stochastic variable in \citet{2021Park} uses observations from before  MJD$=59200$ (December 2020). We mark this date in the figure with the light-blue, dashed line.}
     
	 \label{fig:lc}
\end{figure*}

\section{Observations}\label{sec:obs}

This section describes the multi-epoch imaging surveys optical and mid-IR wavelengths that have observed the PR Ori system, initial follow-up spectroscopy with IRTF, and subsequent follow-up photometry and spectroscopy.  The photometry is described in Table \ref{tab:phot}. The light curves arising from the resolved and unresolved photometry of the system are shown in Fig. \ref{fig:lc}.

 This work was initiated after identifying high-amplitude variability from PR Ori in mid-IR photometry from all-sky observations of the {\it WISE} telescope \citep{2021Park}. 
 A subsequent spectroscopic follow-up programme, developed to understand the mechanism driving the variability in these objects, has led to the discovery of new eruptive YSOs \citep{2023Contreras_a,2024Ashraf}. The YSO PR Ori was included as a stochastic mid-IR variable, with a NEOWISE W2 amplitude of 1.1 mag \citep{2021Park}.  The mid-IR light curve covered observations with MJD$<\sim59200$ (before December 2020) and therefore did not show the rising outburst, so our follow-up was investigating the cause of PR Ori's stochastic variability and not because it was considered as a potential FUor object.  The spectrum appeared similar to FU\,Ori outbursts, and the outburst was then confirmed with subsequent data releases from NEOWISE and through optical photometry from {\it Gaia} \citep[{\it Gaia} science alert Gaia21arx][]{2021Hodgkin}.

\begin{table*}
\caption{Optical to mid-IR photometry of PR Ori. The full version of the table is available online.}
\label{tab:phot}
  
\resizebox{\textwidth}{!}{\begin{tabular}{lcccc|lcccc|lcccc} 
		\hline

MJD & Mag & Mag error & Filter &  Source & MJD & Mag & Mag error &Filter &  Source & MJD & Mag & Mag error &Filter &  Source\\
\hline
 &  & &  &  & &  &  & & & & \\
\multicolumn{5}{l|}{PR Ori A} & \multicolumn{5}{l|}{PR Ori B} & \multicolumn{5}{l}{PR Ori A+B}\\
 &  & &  &  & &  &  & & & & \\
 \hline
51874.85730 & 9.09 & 0.02 & J & 2MASS$^{a}$ & 51874.85730 & 11.16 & 0.02 & J & 2MASS$^{a}$ & 56004.32362 & 11.49 & 0.01 & V & ASAS-SN\\
51874.85730 & 8.55 & 0.01 & H & 2MASS$^{a}$ & 51874.85730 & 10.25 & 0.01 & H & 2MASS$^{a}$ & 56004.32539 & 11.49 & 0.01 & V & ASAS-SN\\
51874.85730 & 8.39 & 0.01 & K$_{\rm s}$ & 2MASS$^{a}$ & 51874.85730 & 9.85 & 0.01 & K$_{\rm s}$ & 2MASS$^{a}$ & 56016.26176 & 11.47 & 0.01 & V & ASAS-SN\\
55936.42679 & 8.70 & 0.04 & J & R18 & 55936.42679 & 11.31 & 0.04 & J & R18 & 56016.26345 & 11.49 & 0.01 & V & ASAS-SN\\
55936.42679 & 8.48 & 0.04 & H & R18 & 55936.42679 & 10.69 & 0.04 & H & R18 & 56226.53686 & 11.52 & 0.01 & V & ASAS-SN\\
55936.42679 & 8.38 & 0.10 & K & R18 & 55936.42679 & 10.13 & 0.10 & K & R18 & 56226.53861 & 11.51 & 0.01 & V & ASAS-SN\\
59985.26514 & 8.48 & 0.01 & K$_{\rm s}$ & SpeX & 59985.26514 & 9.47 & 0.01 & J & SpeX & 56228.50443 & 11.51 & 0.01 & V & ASAS-SN\\
61008.56416 & 8.30 & 0.01 & K$_{\rm s}$ & SpeX & 59985.26514 & 8.49 & 0.01 & H & SpeX & 56228.50612 & 11.52 & 0.01 & V & ASAS-SN\\
59275.74757 & 12.67 & 0.04 & B & RC80 & 59985.26514 & 7.71 & 0.01 & K$_{\rm s}$ & SpeX & 56230.49368 & 11.55 & 0.01 & V & ASAS-SN\\
59275.75025 & 12.03 & 0.04 & g & RC80 & 61008.56416 & 9.07 & 0.01 & J & SpeX & 56230.49534 & 11.56 & 0.01 & V & ASAS-SN\\
59275.74905 & 11.50 & 0.03 & V & RC80 & 61008.56416 & 8.15 & 0.01 & H & SpeX & 56232.56040 & 11.53 & 0.01 & V & ASAS-SN\\
59275.75103 & 11.12 & 0.02 & r & RC80 & 61008.56416 & 7.31 & 0.01 & K$_{\rm s}$ & SpeX & 56232.56211 & 11.53 & 0.01 & V & ASAS-SN\\
56915.93611 & 11.20 & 0.01 & G &{\it Gaia} DR3 & 59275.74757 & 13.80 & 0.05 & B & RC80 & 56235.51094 & 11.52 & 0.01 & V & ASAS-SN\\
56951.07373 & 11.12 & 0.01 & G &{\it Gaia} DR3 & 59275.75025 & 13.30 & 0.04 & g & RC80 & 56235.51268 & 11.52 & 0.01 & V & ASAS-SN\\
56951.24989 & 11.11 & 0.01 & G & {\it Gaia} DR3 & 57071.74660 & 13.87 & 0.01 & G &{\it Gaia} Alerts & 56237.54377 & 11.51 & 0.01 & V & ASAS-SN\\
57071.74677 & 11.15 & 0.01 & G & {\it Gaia} DR3 & 57071.82059 & 13.85 & 0.01 & G &{\it Gaia} Alerts & 56237.54557 & 11.52 & 0.01 & V & ASAS-SN\\

\hline
\multicolumn{15}{l}{$a$ From analysis in this work}\\
\multicolumn{15}{l}{R18: Subrau photometry from \citet{2018Reipurth}}\\
	\end{tabular}}
\end{table*}

\subsection{Mid-IR Photometry}

{\it WISE} surveyed the entire sky in four bands, W1 (3.4 $\mu$m), W2 (4.6 $\mu$m), W3 (12 $\mu$m), and W4 (22 $\mu$m), with spatial resolution of $6\farcs1$, $6\farcs4$, $6\farcs5$, and 12\arcsec, respectively, from 2010 January to September \citep{2010Wright}. The survey continued as the NEOWISE Post-Cryogenic Mission, using only the W1 and W2 bands, for an additional 4 months \citep{2011Mainzer}. In September 2013, WISE was reactivated as the NEOWISE-reactivation mission \citep[NEOWISE-R,][]{2014Mainzer}. NEOWISE-R was decommissioned in August 2024, with the final data release containing observations until 31 July 2024. For each visit to a particular area of the sky, {\it WISE} performed several photometric observations over a period of $\sim$few days. Each area of the sky was observed similarly every $\sim$ 6 months.

For the analysis of PR Ori, the single-visit data were collected from the NASA/IPAC Infrared Science Archive (IRSA) catalogues using a 3\arcsec\ search radius from the coordinates of the YSO. In addition to the single-visit data, we also construct a light curve where photometry taken over a period of a few days is averaged to produce one epoch of photometry every six months \citep[following section 4.1 in ][]{2021Moor}. We also applied saturation photometric bias correction as described in Section II.1.c.iv.a. of the NEOWISE Explanatory Supplement\footnote{\url{https://irsa.ipac.caltech.edu/data/WISE/docs/release/NEOWISE/expsup/sec2_1civa.html}}.

\subsection{Optical photometry}

We obtained optical photometric observations of the PR Ori system with the Astrosysteme ASA800 80\,cm Ritchey-Chr\'etien telescope (RC80) at the Piszk\'estet\H{o} Mountain Station of Konkoly Observatory (Hungary). The RC80 telescope is equipped with an FLI Microline PL230 CCD camera, $0\farcs55$ pixel scale, $18\farcm8\times18\farcm8$ field of view (FoV), Johnson $BV$ and Sloan $g'r'i'$ filters. We obtained about 50 epochs between March 2021 and February 2026, with uneven cadence. On each observing night, we obtained three images of 60\,s exposure time in each filter, which were reduced following the usual steps of bias and dark subtraction and flatfield correction. Due to the proximity of the PR Ori A and B components, we obtained PSF photometry for them and for 14 comparison stars in the field of view,  selected from the APASS9 catalog \citep{apass9} to cover a sufficiently large range in $V-r'$ color.
These comparison stars were used to build the PSF in each image to establish color-dependent zero points from their APASS9 magnitudes.
Typical photometric uncertainties are 0.05\,mag in $B$, 0.04\,mag in $g'V$, and 0.03\,mag in $r'i'$.

\subsection{Archival optical and near-IR photometry}\label{ssec:opt}

The Zwicky Transient Facility \citep[ZTF,][]{2019Bellm} is a Northern-equatorial sky survey that performs optical ($g, r, i$) observations of the Northern sky ($\delta>-30~\deg$) with a three-night cadence. The latest data release (DR24) contains observations from March 2018 until March 2026. We queried the IRSA for detections within 3\arcsec\ of PR Ori. No light curves could be retrieved, as the system is brighter than the saturation limit of ZTF (12.5--13 mag). To investigate the reflection nebula around PR Ori (see Section \ref{ssec:optical_outburst}), we collected r-band images of the source using the image query service at IRSA.

We collected $B, V, R, I$ photometry from the American Association of Variable Star Observers (AAVSO) database\footnote{www.aavso.org}. The data covers the period from December 2005 until February 2026. We also include the All-Sky Automated Survey for Supernovae (ASAS-SN) project \citep{2017Kochanek} $V$-band photometry, covering September 2012 to February 2018, and $g$-band photometry, covering September 2017 to February 2025. Finally, we collected ASAS3 \citep{1997Pojmanski} $V$-band photometry covering the period from November 2000 to December 2009.
 
The observations from ZTF, ASAS-SN, ASAS3 and AAVSO do not have enough angular resolution to distinguish PR Ori A and B.

The {\it Gaia} telescope resolves the emission from the A and B components. The epoch photometry of PR Ori A, for the period 2014-Sep-15 to 2017-Apr-30, is found in the {\it Gaia} DR3 release \citep{2023GaiaDR3}. The photometry of PR Ori B was published by the {\it Gaia} Science Alerts \citep{2021Hodgkin} as Gaia21arx, covering the duration of the full {\it Gaia} mission, from 2015-Feb-18 until 2024-Jan-21, showing a 3--4 mag brightening event in the star.

We also measure new 2MASS photometry for PR Ori A and B, since the 2MASS point source catalog did not separate the two components.  The brightness is calculated using PSF photometry on the two components, with a separation fixed by {\it Gaia} astrometry.  The magnitude in both components is then calculated by comparing the brightness on the image to 20 other objects in the field. The photometry and observation date are shown in Table \ref{tab:phot}.


\subsection{IRTF/SpeX Spectroscopy}

We obtained near-IR spectra of PR Ori A and B  using SpeX \citep{rayner03} mounted at the NASA Infrared Telescope Facility (IRTF) on Mauna Kea (programmes 2023B079 and 2025B096, PI Contreras Pe\~{n}a, and 2023A035, PI K\'osp\'al). PR Ori A was observed on 29 November 2025 (UTC) using the short wavelength cross-dispersed (SXD) mode (0.7--2.5 $\mu$m). The total integration time of the observation was 240s, using the $0\farcs8$ slit ($R=750$).

PR Ori B was observed on four separate occasions, on 10 February 2023, 23 December 2023, 20 August 2025 and 29 November 2025 (UTC), using the SXD mode. The source was also observed using the long wavelength cross-dispersed (LXD) mode during the February 2023 observations. The $0\farcs8$ slit ($R=750$) was used in the February 2023 and November 2025 observations, while the $0\farcs5$ slit ($R=1200$) was used for the December 2023 and August 2025 observations. The total integration times were 600s (SXD, February 2023), 200s (LXD, February 2023) , 300s (SXD, October 2023), 720s (SXD, December 2023), 240s (SXD, August 2025) and 360s (SXD, November 2025). Bright A0V standard stars were also observed for telluric calibration (HIP 26812, HIP 24607, and HIP 18769). All spectra were reduced and calibrated using Spextool version 4.1 \citep{cushing04}.


For the SpeX observations of 10 February 2023, we obtained photometry from K-band images with the guide camera for PR Ori B and HD 37131. The flux calibration from the telluric spectra and from the guide camera photometry agree within 0.037 mag (a difference less than 4\%), suggesting that the night was stable and the flux calibration is robust, within 5--10\%. Eventually, the spectra were rescaled so that they reproduce the K-band photometry with the guide camera. We calculated synthetic photometry from the SpeX spectra of PR Ori B using the 2MASS JHK filter profiles. Using the guide camera images we also obtained K$_{\rm s}$ photometry for PR Ori A. The photometric data is presented in Table \ref{tab:phot}.

\subsection{BOAO/BOES}

We observed the PR Ori system using the Bohyunsan Optical Echelle Spectrograph (BOES) on the 1.8 m optical telescope at Bohyunsan Optical Astronomy Observatory (BOAO) in the Republic of Korea. The observations, obtained in November 2024 and February 2025, cover from 3600--10500 \AA\ with a resolving power of 45000 using the 200$\mu$m fiber.  The spectra were reduced with the IRAF echelle package. Each aperture from the spectral images was extracted using a master flatfield image. Using the flatfielding process, we corrected the interference fringes and pixel-to-pixel variations of the spectrum images. A ThAr lamp spectrum was used for wavelength calibration.

\subsection{Palomar/NGPS Optical Spectroscopy}

We obtained optical spectra of PR Ori A and B on 2025 October 27 with NGPS \citep{jiang_ngps} during commissioning on the Hale 200-inch Telescope at Palomar Observatory. Our observations cover the r and i bands, yielding spectra spanning 5600--13000 \AA, with $R\sim4000$.  The u and g-bands were not yet available at the time of the observations.  
The observations used a $0\farcs5$ slit width with the position angle set to 305$^\circ$.   The total integration time was 300 s.  

 Data were reduced following standard procedures for long-slit spectroscopic data reduction using a custom pipeline developed for NGPS. Wavelength calibration was performed against daytime ThAr and FeAr arc exposures. Flux calibration was performed with an observation of the spectrophotometric standard G191-B2B taken immediately following the science observation.

\subsection{GALAH DR4}

 PR Ori B was observed by the Galactic Archaeology with HERMES (GALAH) Survey \citep{2025Buder}. The observations were carried out on 2022 February 2 using the HERMES spectrograph (R$\sim$30000) on the 3.9 m Anglo Australian Telescope. The instrument is a fiber-fed spectrograph covering four non-contiguous bands covering between 4700 and 7900 ${\rm \AA}$. The projected slit width is 2", and the coordinates of the observations are centered on PR Ori B. In this work we make use of the single co-added spectrum of the YSO provided in the data release 4 of the survey.

\subsection{VLT/VISIR mid-IR spectroscopy}

We obtained mid-IR spectroscopy of PR Ori B using the ESO/VLT VISIR spectrograph (112.25YV, PI: P. \'Abrah\'am). The observation was carried out on 2023 November 4 at an airmass of 1.06. We used the low resolution spectroscopy mode with a 1$''$ wide slit, providing a spectral resolution of $R \sim 350$ in the 8--13$\,\mu$m  range. The observing sequence followed an ABBA nodding pattern along the slit. To ensure precise telluric correction and flux calibration, the K5III star HD\,48217 was observed before and after PR Ori B, and the spectral response function was interpolated for the epoch of the science observation. Basic data reduction and extraction of the 1D spectrum was performed with the ESO VISIR spectroscopic pipeline Version 4.4.4 in the EsoRex environment. Flux calibration was performed by multiplying with the model spectrum of HD\,48217.

\subsection{Observations from \citet{2018Reipurth}}

We also present the optical and near-IR spectroscopic data from \citet{2018Reipurth}. The near-IR observations were carried out using GNIRS at Gemini-North during 2012 February 4 for PR Ori A, and 2013 January 21 for PR Ori B, and with a resolving power of $R\sim1700$ and $R\sim800$, respectively.

\citet{2018Reipurth} obtained Keck/HIRES echelle spectra of PR Ori A and B on 2010 Nov 14 and 2011 Dec 18, spanning 4700-8700 \AA, with seeing of $\sim 0\farcs9$ and $\sim1\farcs1$, respectively.  The slit was aligned with the binary, so both components are well separated on the detector.
We reduced and reanalyzed the spectra of both components.
The spectra were extracted by simultaneously fitting two Gaussian profiles of the same angular width in the cross-dispersion direction and separately with box extractions, with spectra consistent with the figures in \citet{2018Reipurth}.  Some residual OH and O$_2$ night sky emission lines are present in the reductions and are subtracted by visual inspection.
The flux of PR Ori B is calibrated by assuming that PR Ori A is constant, as shown by the {\it Gaia} lightcurve, and then adopting the {\it Gaia} spectrum.  The 2010 November 14 data was obtained in better weather, which led to higher S/N and a better sky subtraction. 

\subsection{ALMA Observations} \label{sec:Obs_datareduction}

The PR Ori system was observed on 2024 August 1 (2023.1.00561.S, PI: Vioque, Miguel).  The observations were originally published in a large survey of Herbig AeBe stars by \citep{stapper25}  One execution block failed during restoration of the delivered final flag version because of a row-number mismatch in the flag table, so the analysis here uses the three successfully restored execution blocks.

We used the continuum and $^{12}$CO product data from the ALMA archive to constrain the disk properties of PR Ori B. For the continuum fit, the spectral windows centered on the $^{12}$CO line near 230.538 GHz were excluded, and the remaining continuum spectral windows were averaged to produce a compact visibility table.  The synthesized beam size of the continuum data is 0\farcs36$\times$0\farcs28, with position angle (PA) of 85.56$\degr$, while for the $^{12}$CO data we estimate a beam size of 0\farcs37$\times$0\farcs29, and PA of 81.98\degr. The root-mean-square (rms) values are 4.04$\times$10$^{-2}$ mJy beam$^{-1}$ ($\sigma_{\mathrm{cont}}$) and 4.08 mJy beam$^{-1}$ ($\sigma_{^{12}\mathrm{CO}}$) for the continuum and $^{12}$CO data, respectively.

\section{Before the eruption: A Historical overview of the PR Ori system}\label{sec:hist}

The PR Ori system is one of the brightest members of the L1641 region.
In a dedicated study of the PR Ori system, \citet{2018Reipurth} found that it is comprised of PR Ori A, itself a close ($0\farcs077$) binary, and PR Ori B (Gaia21arx, ESO-H$\alpha$ 1481, $\alpha =$05:36:24.8, $\delta =-$06:17:31), separated by $3\farcs5$\footnote{A fainter source, PR Ori C, is located $\sim 10$\arcsec\ away and was thought to be a distant companion, but the {\it Gaia} proper motion indicates that the object is either a field star or a runaway star from some other system in Orion.}.  Excess infrared emission and sub-mm emission, both produced by a circumstellar disk, are associated with PR Ori B but not with PR Ori A \citep{2013Fang,2018Reipurth,2022vanTerwisga}.

The PR Ori system was previously recognized as a variable star\footnote{Reported along with complaints about funding!}, with observations dating to 1888 that show brightness variability in the visible range from 13.0--14.5 mag \citep[source 64,][]{1904Pickering}. Comprehensive monitoring described by \citet{1954Parenago}, including epochs dating to 1864, revealed similar variability of 1--1.5 mag, and with more faint epochs than bright epochs.

\begin{figure}
\centering
\resizebox{\columnwidth}{!}{\includegraphics[angle=0]{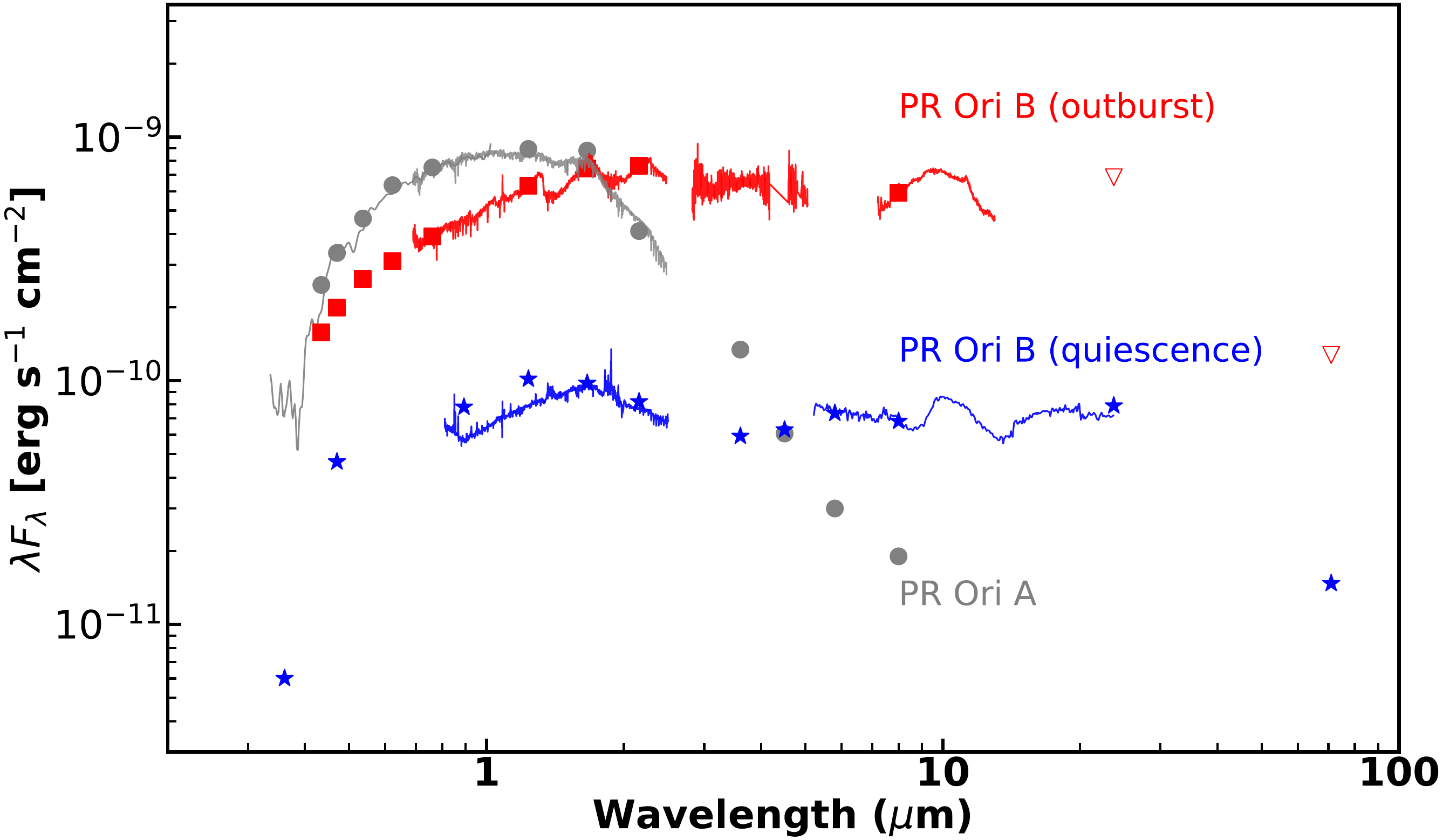}}
	 \caption{ Spectral energy distribution of PR Ori A (grey) and PR Ori B during quiescence (blue) and outburst (red). The flux is determined from photometry (circles, squares and stars) and spectra (solid lines). For $\lambda>10~\mu$m, the outburst flux is derived by assuming that the change is similar to the increase observed at 8~$\mu$m.  In the figure, downward triangles mark upper limits of the outburst flux at $\lambda>10~\mu$m.}
	 \label{fig:sed}
\end{figure}

PR Ori B is also the driver of the Herbig-Haro object HH305, with a series of shocks that may indicate past outbursts \citep{2018Reipurth}.
When searching for nebulous features around star forming regions using Schmidt plates, \citet{1985Reipurth} identified this group of shocks that lie along a line passing through PR Ori \citep[]{1998Reipurth,1999Mader}.

The small near-infrared (near-IR) excess and weak H$\alpha$ emission of PR Ori \citep{1990Strom} and the offset of the location of the primary with respect to some of the knots of HH305 \citep{1998Reipurth}, indicated that a fainter secondary source is the likely driver of HH305. Support for this scenario is provided by the identification of PR Ori as an optical double in \citet{1990Strom} and the formal identification of PR Ori B by \citet{1999Mader}.  PR Ori B is classified as an H$\alpha$ emission line star, though the emission strength is not particularly high \citep[ESO-H$\alpha$ 1481,][]{2014Pettersson}.  Although historically PR Ori A is brighter than PR Ori B at optical and near-IR wavelengths, PR Ori B in quiescence dominates emission in the mid-IR and at longer wavelengths \citep{2012Megeath, 2016Chen}. 

The most in-depth investigation of PR Ori B is provided by \citet{2018Reipurth}, with an extensive analysis of optical and near-IR spectra and the spectral energy distribution (SED) from the optical to the far-IR.
PR Ori B has a near-IR spectral type of K7$\pm$2 and shows H$\alpha$ emission with EW$=-16.8\pm0.1$\,\AA, with a blueshifted absorption dip ($v=-90$\ km s$^{-1}$) from a wind, and a P\,Cygni profile in \ion{He}{1} $\lambda$10830.  

\begin{figure*}
\resizebox{1.3\columnwidth}{!}{\includegraphics[angle=0]{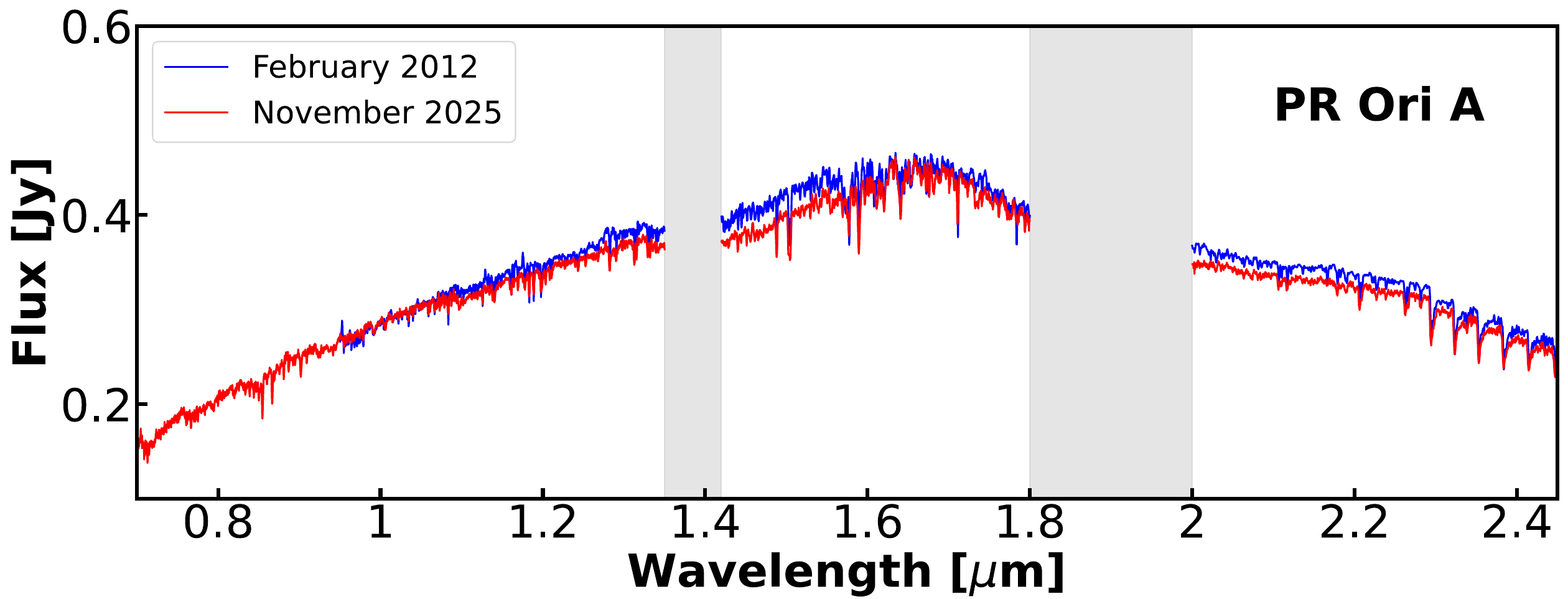}}
\resizebox{0.7\columnwidth}{!}{\includegraphics[angle=0]{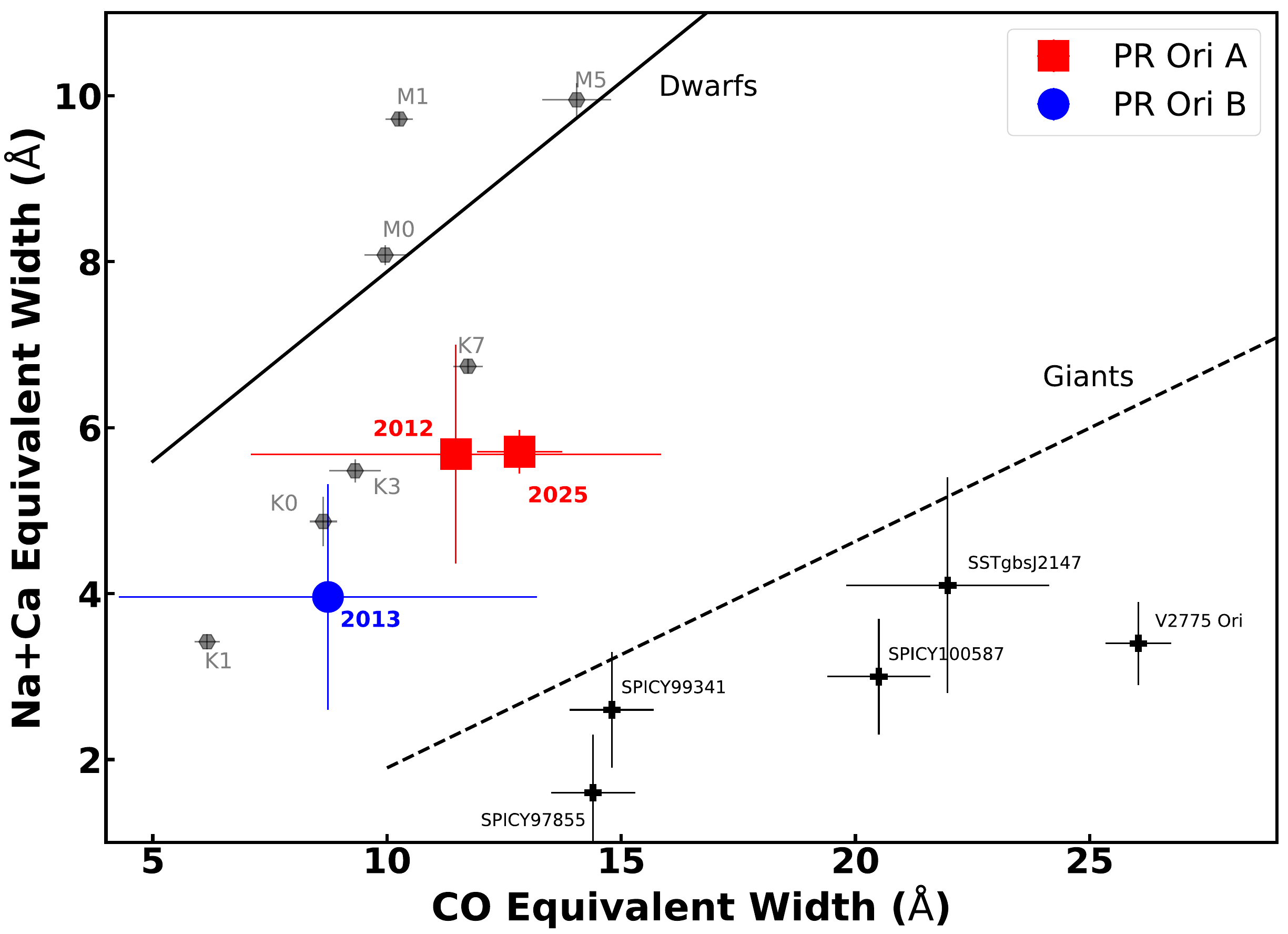}}
	 \caption{(Left) Near-IR spectrum of PR Ori A at different epochs. The grey-shaded areas mark regions strongly affected by telluric lines. (Right) Equivalent Width of Na I$+$Ca I versus $^{12}$CO for PR Ori A (red squares) and PR Ori B (quiescence, blue circle) measured at different dates (marked in the figure). The values for a sample of known FUors (see Table \ref{tab:app}) are also presented in the figure. The expected values for main-sequence and giant stars are shown by black solid and dashed lines, respectively.}
	 \label{fig:ew}
\end{figure*}

\begin{figure*}
\resizebox{\textwidth}{!}{\includegraphics[]{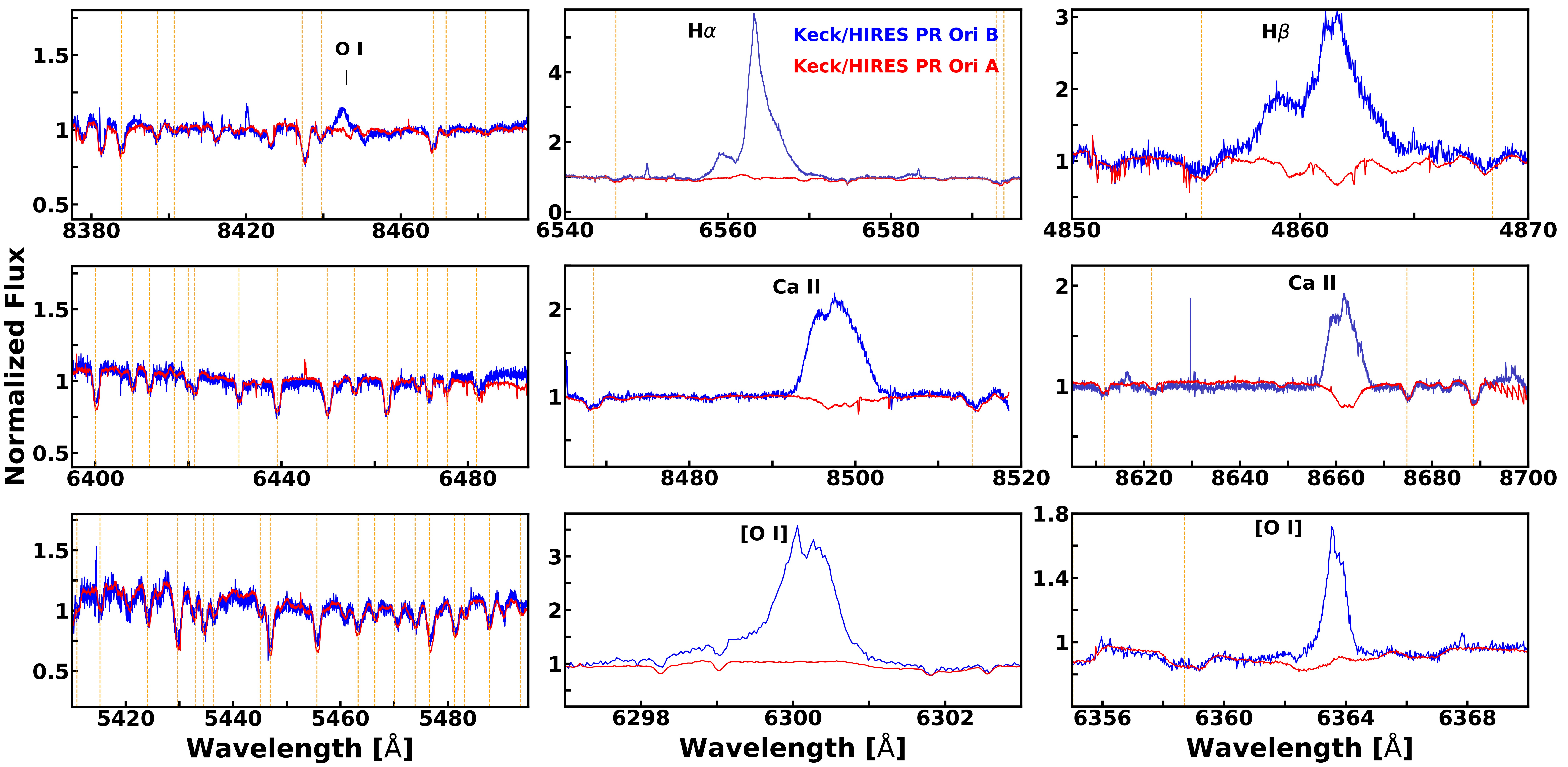}}
	 \caption{November 2010 Keck/HIRES spectra of PR Ori A (red) and PR Ori B (blue), both normalized at the middle of the order and binned by 3 pixels.  The photospheric absorption lines (marked by the dotted orange lines) of PR Ori A match exactly those of PR Ori B. Some emission lines, marked in the panels, are also present in the PR Ori B spectrum.}
	 \label{fig:hiresspecs}
\end{figure*}

\subsection{The Properties of the Progenitor}

PR Ori B is classified as a K7 star from the near-IR spectrum \citep{2018Reipurth}. This corresponds to a star with an effective temperature of T$_{\rm eff}=4050$~K \citep{2013Pecaut}. Using the distance of 388.4\,pc, extinction A$_{V}=3.1$~mag (adopted from the outburst measurement), and a bolometric correction of 1~mag \citep{2013Pecaut}, we calculate a quiescent stellar luminosity of 1.27\, L$_{\odot}$ from the H-band absolute magnitude of PR Ori B during quiescence. The luminosity and effective temperature correspond to a stellar radius of R$_{\ast}=2.3$~R$_{\odot}$. From the models of \cite{2020Somers} with 34\% spot coverage, we estimate a mass of M$_{\ast}=0.85$~M$_{\odot}$ and an age of 1.23 Myr. 

We caution that these values are derived by adopting the $A_V$ from the outburst and assuming that the $A_V$ remained constant with time. However, some changes may have occurred between quiescence and outburst, either decreasing $A_V$ via dust clearing \citep[as observed in V1647 Ori][]{2011Aspin} or increasing $A_V$ due to dusty winds \citep[which may play a role in HBC722][]{herczeg2025}. In addition, the spot coverage of young K-type stars could be as high as 80\% \citep{2025Perez}.


 Figure \ref{fig:sed} shows the SEDs of the A and B components of the PR Ori binary system, including quiescence and outburst for B and the {\it Gaia} XP and IRTF/SXD spectra of PR Ori A.  For PR Ori B we show SEDs and spectra taken during quiescence and outburst, including the Spitzer/IRS observations from 2007 (quiescence from \citealt{2016Kim}) and the IRTF/SXD, IRTF/LXD and VLT/VISIR spectra observed during the outburst.

The SED of the PR Ori system shows that PR Ori B during quiescence has far more
circumstellar material than the A component.  We use the SEDs between 2 and 20 $\mu$m to determine the spectral index, $\alpha$, of the A and B components. For PR Ori A we determine $\alpha=-2.5$, consistent with a Class III YSO, while for PR Ori B (during quiescence) we obtain $\alpha=0.03$, or a Flat-spectrum YSO \citep[following the prescriptions of][]{gutermuth09}.



\subsection{The stellar properties of PR Ori A}

The stellar properties of PR Ori A, as measured from high-resolution Keck I/HIRES spectra, indicate that the system is young.  PR Ori A itself is an intermediate mass T Tauri star and precursor to an A star.
We measure stellar properties of PR Ori A by first measuring the stellar temperature from the Keck I/HIRES spectrum with the SpecMatch-EMP pipeline \citep{2017Yee}.  Orders with good fits yield  $T_{\text{eff}} = 4940 \pm 110\, \mathrm{K}$, and $v \sin i = 52 \pm 7 \, \mathrm{km \, s^{-1}}$.  Our own $\chi^2$ minimization fits of BT-Settl (AGSS2009) spectra to the observed spectrum yield similar results, with $T_{\text{eff}} = 5100 \pm 400 \, \mathrm{K}$ and $v \sin i = 47 \pm 4\, \mathrm{km \, s^{-1}}$, and radial velocity $26.8 \pm 0.7 \, \mathrm{km \, s^{-1}}$.
This temperature is consistent with the K2 spectral type assessed by \citet{2018Reipurth} and G9 from \citet{2013Fang}.

The equivalent widths of \ion{Na}{1}$+$\ion{Ca}{1} versus $^{12}$CO of our 2025 near-IR spectrum of PR Ori A are also consistent with a mid-K spectral type (see Fig.\ \ref{fig:ew}). The similarities between the observations by \citet{2018Reipurth} and our spectrum (Fig. \ref{fig:ew}), taken $\sim$11 years apart, provides further evidence that PR Ori A is not the source driving the large variability in the system.

Adopting the $4940$\,K temperature, we obtain $A_V=0.73$ mag from the {\it Gaia} colors and \citet{2013Pecaut} temperature-color relationships.  The corresponding luminosity is then 7.8\,$L_\odot$.  We convert the temperature and luminosity to a mass of 1.84\, $M_\odot$ and age $0.63$\,Myr using the Dartmouth pre-main sequence evolutionary tracks \citep{2008Dotter}.

\subsection{Optical Emission from PR Ori B: scattered light from PR Ori A}


During quiescence the photospheric absorption lines of PR Ori B are an exact match for PR Ori A.  Figure~\ref{fig:hiresspecs} shows high-resolution spectra of PR Ori A and PR Ori B from three different spectral ranges.  The lines, line depths, centroid velocity, and rotational velocity of PR Ori B are the same as those of PR Ori A.  
The only significant differences between the spectra are the presence of emission lines from PR Ori B.
The pre-burst optical spectrum from PR Ori B is interpreted as the combination of reflected light from PR Ori A plus some lines produced by accretion processes.

The emission lines detected to PR Ori B include H Balmer and \ion{He}{1} lines, the \ion{Ca}{2} infrared triplet, wind absorption lines, and emission in forbidden lines \citep[see discussion in][]{2018Reipurth}. The emission lines from PR Ori B and its environment sit on top of a photospheric spectrum and are best measured after subtracting the A-component.
These lines are typically associated with accretion and wind processes, so accretion in the system is associated with PR Ori B, as expected from its excess IR emission.  The \ion{Ca}{2} IR triplet lines are bright, as is also seen in the pre-outburst spectrum of the FU\,Ori object HBC 722 \citep{herczeg2025}.


In this interpretation, the strength of the continuum and the line emission from PR Ori B is puzzling.  The emission from PR Ori B is redder than that from PR Ori A, with a flux ratio of 14.5 at 8450 \AA\ and 37 at 5450 \AA\ in 2010, and 14.3 at 8450 \AA\ and 20 at 5450 \AA\ in 2011.
In 2010, these ratios are consistent with an extra $A_V=2.2$ mag to PR Ori B, with dust either between PR Ori A and PR Ori B or in our line of sight to PR Ori B.  The flux ratio in the red is then only  $\sim 5$, after correcting for the excess extinction in the PR Ori B spectrum. The scattering surface must be uncomfortably large and along our line of sight to ensure efficiency by forward scattering emission.  Moreover, the equivalent widths of H$\alpha$ and other lines must be at least a factor of $\sim 10$ larger when calculated 
with respect to the undetected local photosphere of PR Ori B rather than the scattered light continuum.

We therefore cannot reliably use the optical emission to measure any photospheric properties of PR Ori B prior to the FU\,Ori outburst, at least without further processing and analysis.  The effect of scattering on the near-IR emission is unclear, though the near-IR emission from PR Ori B is substantively different from that of PR Ori A \citep{2018Reipurth}.

\subsection{The Disk Properties from ALMA}\label{sec:alma}

\begin{figure}[!b]
    \centering
    \includegraphics[width=0.85\linewidth]{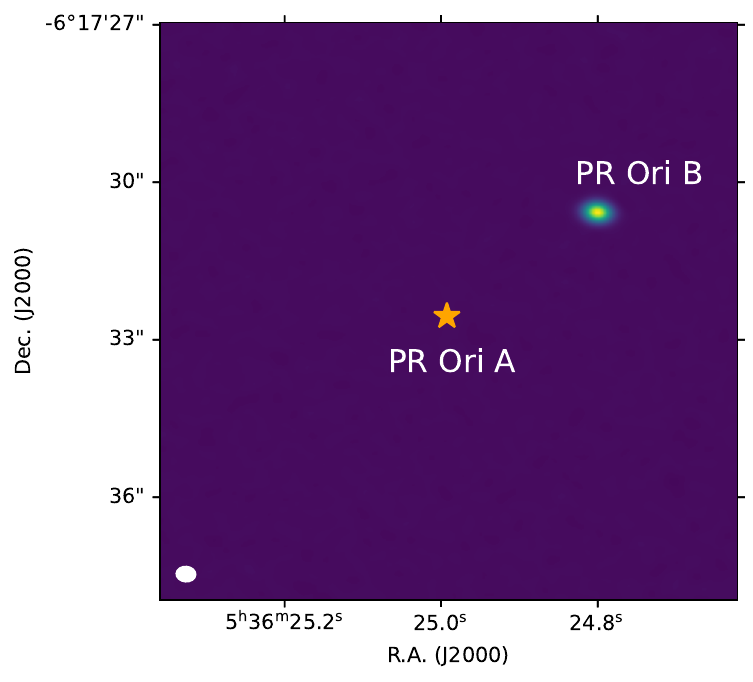}
    \caption{Cleaned image of the non-CO continuum data. The star cross marks the phase center, PR Ori A.  The emission peak is located at the expected offset of PR Ori B.}
    \label{fig:dirty}
\end{figure}

The ALMA 1.3 mm observations were obtained during the outburst, in August 2024.  Dust continuum emission is detected only at the position of PR Ori B (see Fig. \ref{fig:dirty}), while CO emission traces the outflow.


\begin{figure*}[!t]
    \centering
    \includegraphics[width=0.45\textwidth]{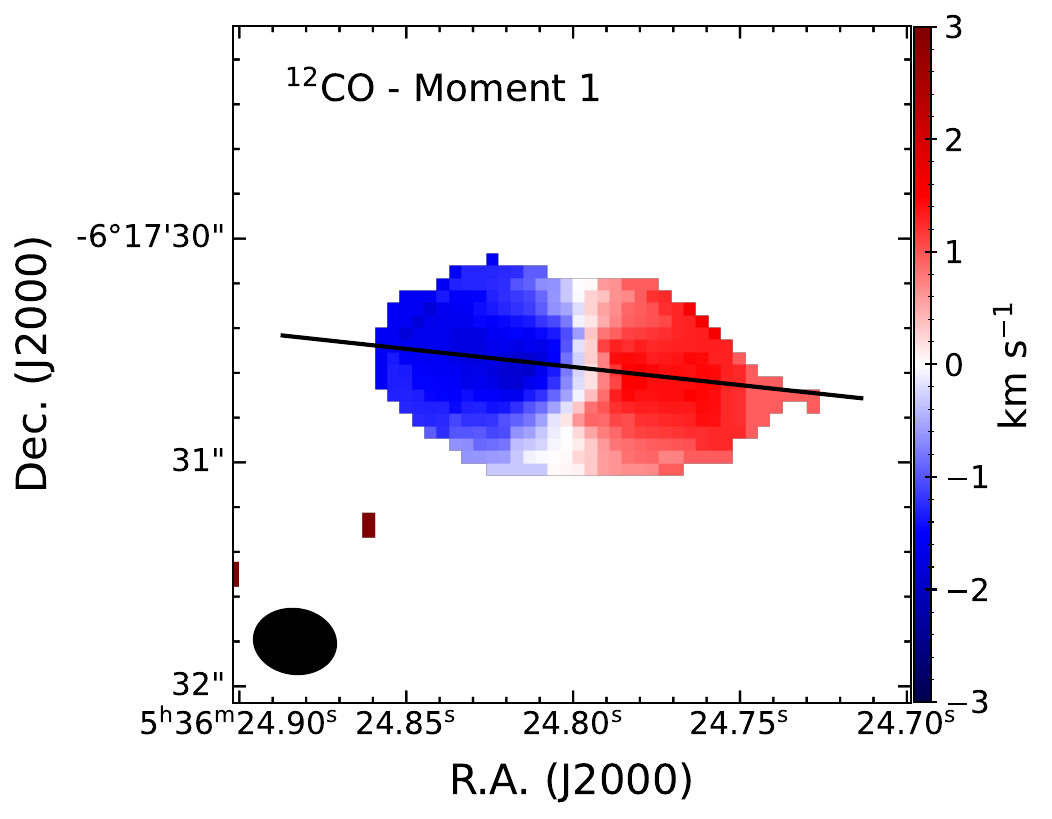}
    \includegraphics[width=0.45\textwidth]{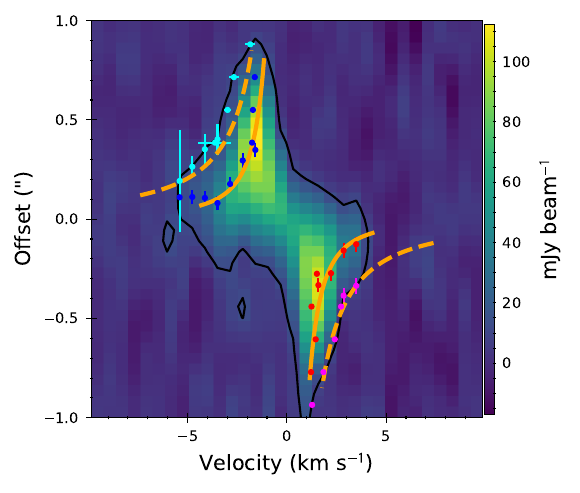}
    \vspace{-0.3cm}
    \caption{\textit{Left}: intensity-weighted velocity (moment 1) map of the $^{12}$CO emission, constructed using emission above 4$\sigma_{^{12}\mathrm{CO}}$ over the velocity range of -10 to 10~\kms\, relative to the systemic velocity ($v_{\mathrm{sys}}$ = 4.6~\kms). The black solid line marks the position-velocity (PV). The synthesized beam is shown in the bottom-left corner. \textit{Right}: PV diagram of the $^{12}$CO. The black contours indicate the 3$\sigma_{^{12}\mathrm{CO}}$ level. The blue and red points show data points extracted using the ridge method, while the cyan and magenta symbols correspond to the points extracted using the edge method implemented in {\tt pvanalysis} within the {\tt SLAM} \citep{Aso2024}. The orange solid and dashed lines show the best-fit results for ridge and edge methods, respectively.}
\label{fig:12co_mom1pv}
\end{figure*}

To quantify the source size more directly, we fit the continuum visibilities with 
an elliptical Gaussian component at PR Ori B, using our own codes that follow methods developed in the Galario program \citep{tazzari18}. 
The fitted Gaussian has a major-axis FWHM of $0.264^{\prime\prime}$ and an axis ratio $q=0.684$, corresponding to a minor-axis FWHM of $0.181^{\prime\prime}$ (approximately $103\times70$ au), at a position angle of $174^\circ$. The total flux from the fit is 24.1 mJy.  The continuum emission is only marginally resolved in the visibility plane, so we conclude that the disk is moderately inclined.
If interpreted as an intrinsically circular, geometrically thin disk, this axis ratio implies an apparent inclination of $i=\cos^{-1}(q)=47^\circ$. The formal statistical uncertainties from the visibility fit are very small and likely underestimate the true systematic uncertainty.  Disks commonly have substructures, including some that are azimuthally asymmetric, which would affect the inclination measurement.  Any envelope contribution at small scales would also affect the size measurements.

The $^{12}$CO 2-1 emission from PR Ori B is seen with velocity shifts consistent with disk rotation (Figure~\ref{fig:12co_mom1pv}).  We analyze the PV diagram using {\tt pvanalysis} in the {\tt SLAM} Python package \citep{Aso2024} to calculate best-fit solutions for the protostellar mass and power-law index of the rotation profile by fitting to the ridge of the emission\footnote{If fitting the edge of the emission, the mass is 2.7\,M$_\odot$ with a power law of 0.68.  However, the fit to the signal centroid should lead to a more accurate mass, and the power law index is higher than expected for a disk.  This edge method leads to a factor of two overestimate in stellar mass \citep{Aso2024}.}. The protostellar mass is 1.01\,M$_\odot$ with a power-law index for the radial emission profile of 0.56, assuming a fixed inclination of $47^\circ$.  This mass is roughly consistent with the mass from the HR diagram, although past measurements have indicated that the method may underestimate masses by 30\% \citep{Aso2024}. 

These fits assume that the dust and gas emission are produced in the disk.  However, a remnant envelope may contribute to both components and would affect our measurements, especially because the dust is only marginally resolved.



\section{The Eruption of PR Ori B}\label{sec:pr_erup}

\begin{figure}[]
\centering
\resizebox{\columnwidth}{!}{\includegraphics[angle=0]{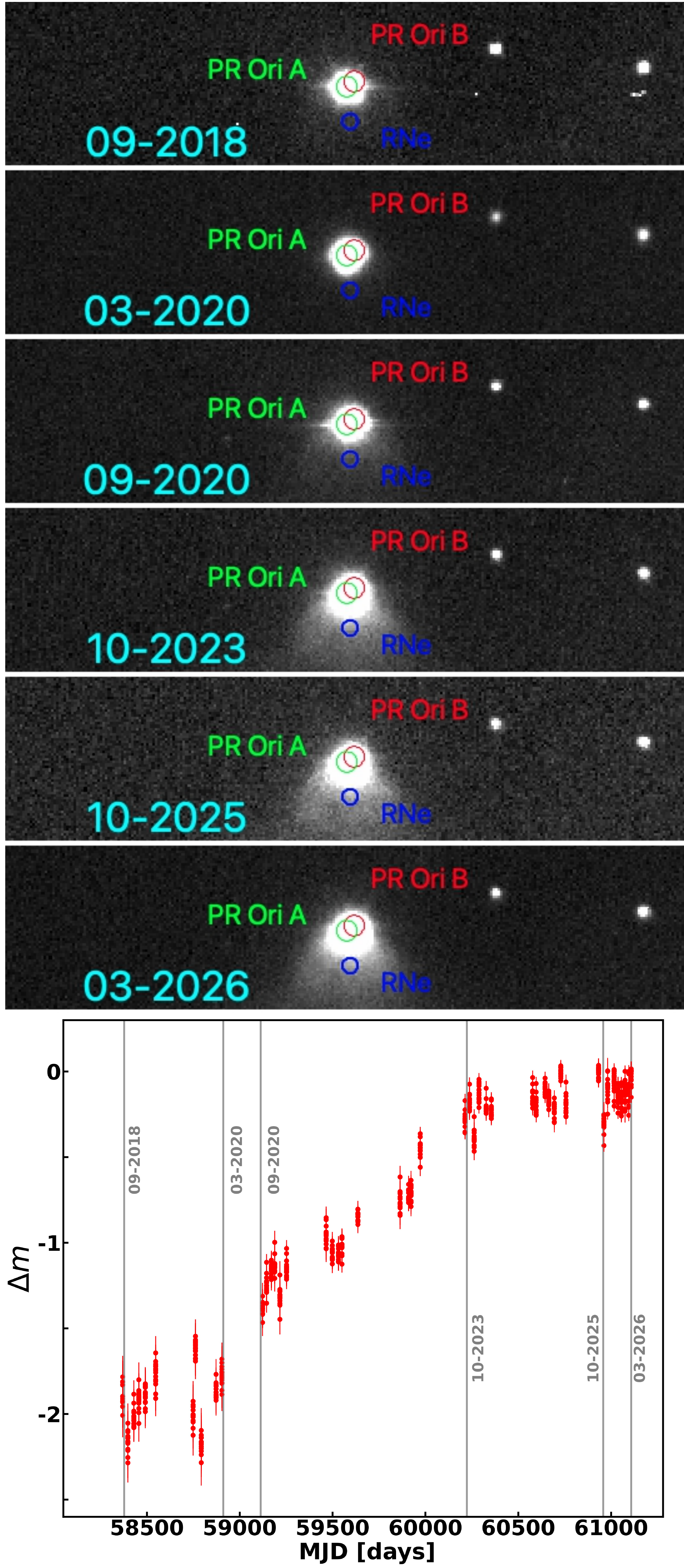}}
\caption{(Top) r-band images from ZTF at different stages of the outburst (with the dates marked in the figures). The blue circle marks the region selected to perform photometry of the nebula. (Bottom) Differential photometry of the nebula against comparison stars around PR Ori (see main text). 
\label{fig:im_ztf}}
\end{figure}

Figure \ref{fig:lc} shows the optical and mid-IR light curves of the PR Ori system.  Over the past decade, PR Ori B has undergone two distinct brightening events, one from MJD 57300--58200 (October 2015 to March 2018) and a second starting from $\sim 58700$ (August 2019) and continuing, as seen in all bands and described in the following subsections.

Some facilities, such as {\it Gaia}, resolve the A and B components, while others, such as ASAS-SN and NEOWISE, provide only total measurements of the system.  
This variability is associated only with PR Ori B.  The primary component, PR Ori A, remains stable at $G=11.15\pm0.04$~mag in Gaia DR3 photometry and in resolved 
 photometry with the RC80 telescope.  
Meanwhile, PR Ori B shows high-amplitude variability in {\it Gaia}, with $\Delta G=5$~mag (Fig.~\ref{fig:lc}) and $\Delta W2\sim2.8$ mag.

For the remainder of this paper, we infer that 
 all episodes of brightening and fading are associated with PR Ori B, with the amplitude dampened by PR Ori A \citep[consistent with the conclusions of][]{2018Reipurth}.  This inference allows us to measure changes attributed to PR Ori B in unresolved photometry, after subtracting off the constant emission from PR Ori A.

\begin{figure*}
\centering
\resizebox{\columnwidth}{!}{\includegraphics[angle=0]{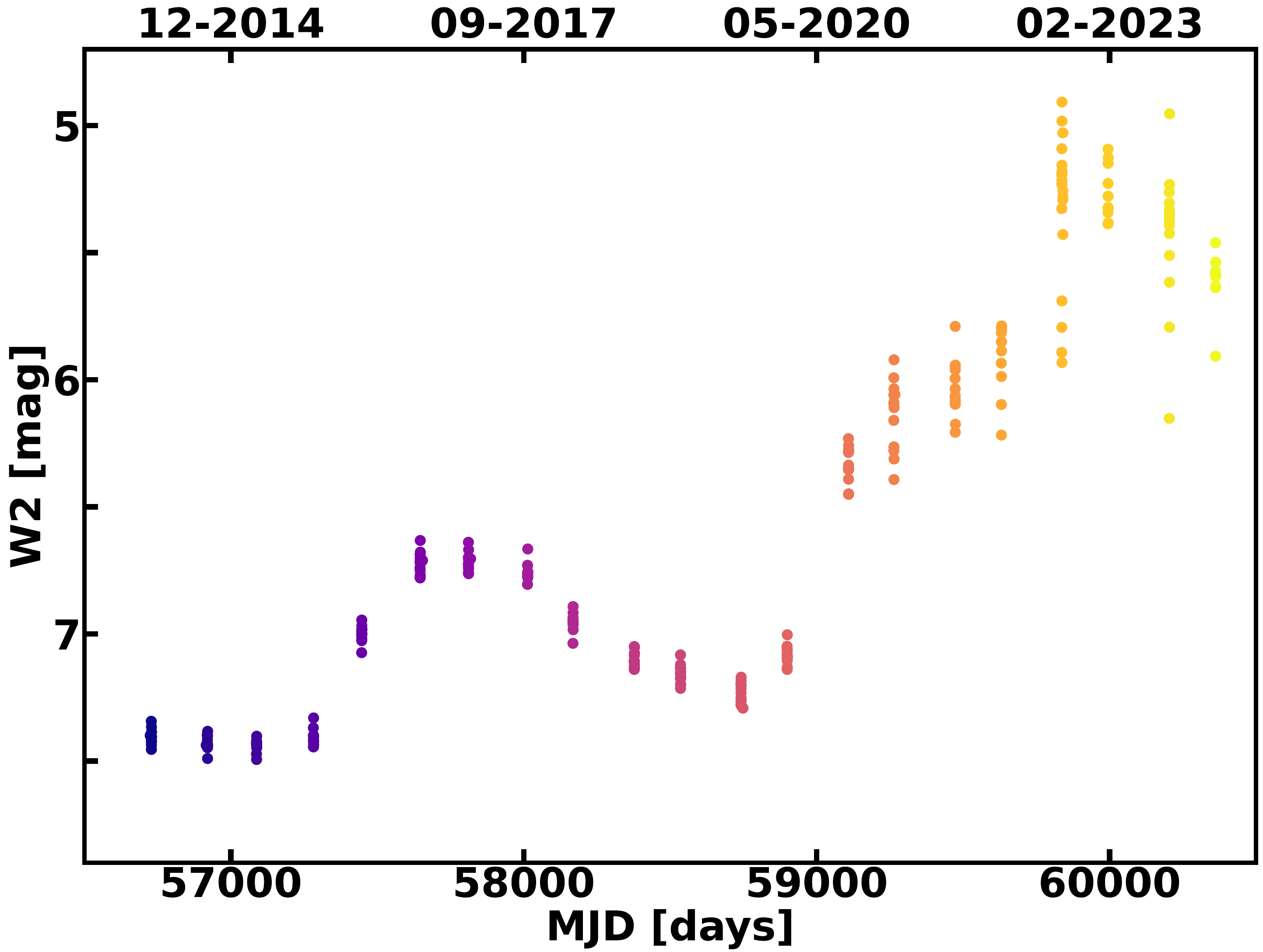}}
\resizebox{0.98\columnwidth}{!}{\includegraphics[angle=0]{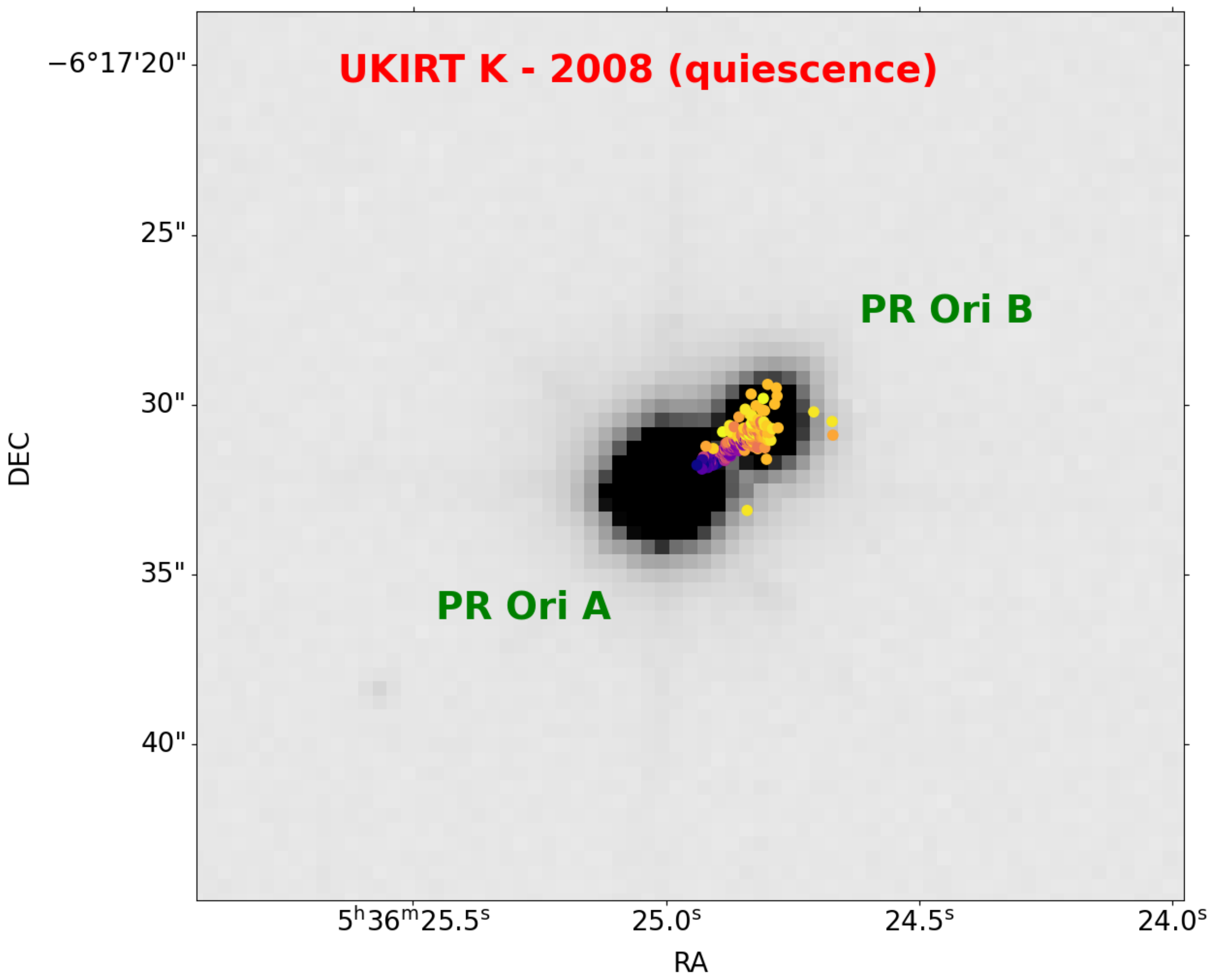}}
\caption{(Left) Single-epoch (colored circles) {\it WISE} W2 photometry of PR Ori. (Right) Right ascension and declination of the single-epoch detections from NEOWISE (colours are the same as the figure in the left). The values of right ascension and declination are placed on top of the UKIRT K-band image of PR Ori (taken in February 2008 during quiescence) for clarity. The source is brighter than the {\it WISE} saturation limit ($W2=6.4$~mag) for detections with MJD$>59000$ (May 2020). This leads to the scatter in the magnitude and location of the single-epoch detections at these dates, even after applying saturation correction. }
\label{fig:mid_ir}
\end{figure*}

\subsection{The outburst in optical photometry}\label{ssec:optical_outburst}

The optical {\it Gaia} photometry shows both brightening events of PR Ori B.  The first brightening occurred at the start of the {\it Gaia} monitoring with an amplitude of $\Delta G=1.6$~mag, although the total brightening could be larger as the outburst might have started prior to the {\it Gaia} coverage.  
After the first brightening, PR Ori B
fades by 3.9 mag and reaches the faintest point of the light curve at MJD$=$58241 (May 2018). 
It then rises by 3.8 mag over 700 days. After reaching $G=12.6$~mag the source continues to rise at a slower rate over the next 1400 days, reaching $G=11.3$~mag in the latest epoch provided by the {\it Gaia} alerts. Within this long-term rise, a fading event, lasting about 600 days and with $\Delta G=0.8$~mag, occurs at MJD$=$59480 (September 2021).

\begin{figure*}[!t]
	\resizebox{\textwidth}{!}{\includegraphics[angle=0]{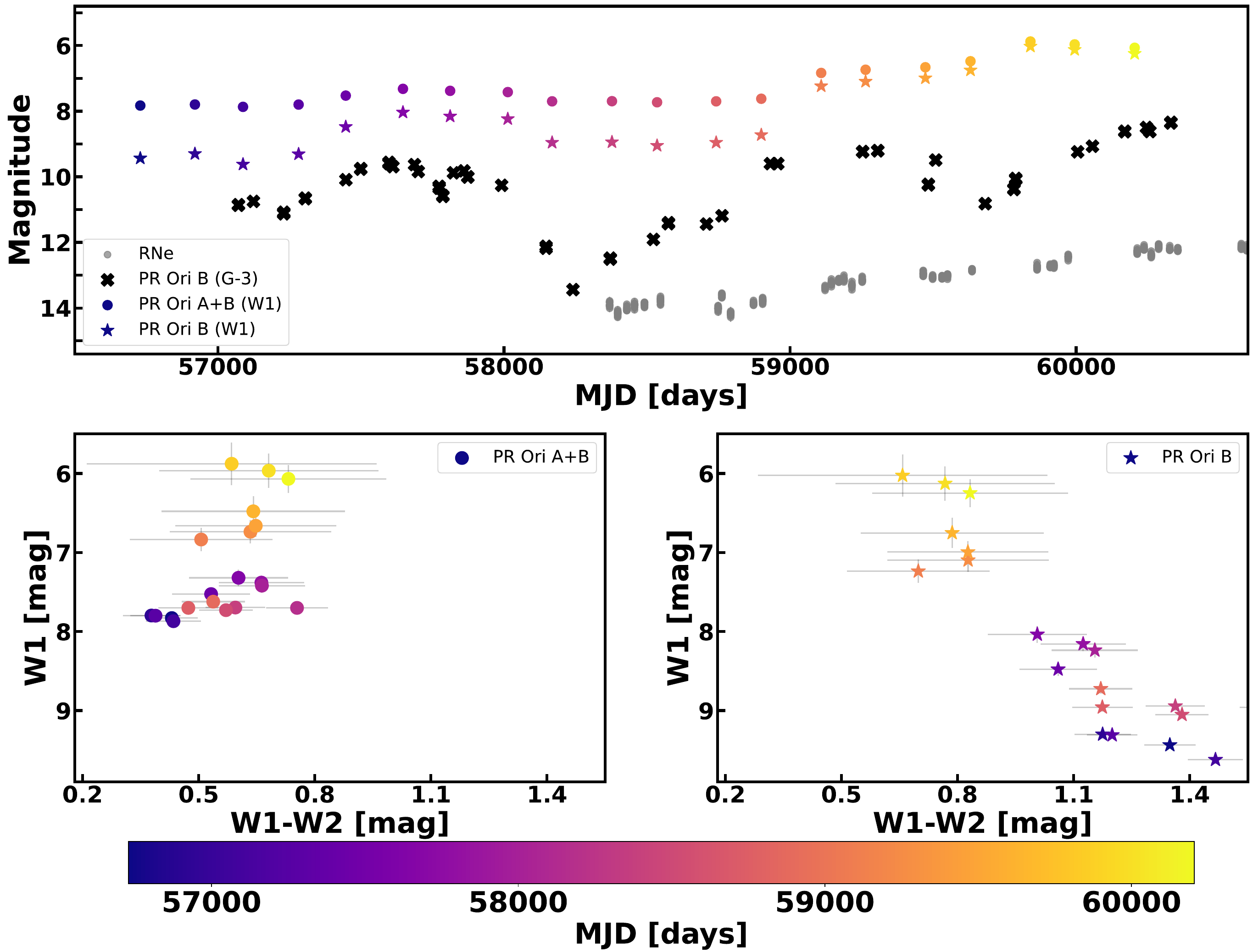}}
	 \caption{(top) {\it Gaia} G photometry of PR Ori B (black), {\it WISE} W1 photometry of PR Ori A+B, and {\it WISE} W1 photometry of PR Ori B (after removing the emission from the A component). The W1 photometry is shown in coloured symbols. (Bottom, left) W1 vs W1$-$W2 diagram for PR Ori A+B. (Bottom, right) W1 vs W1$-$W2 diagram for the emission of PR Ori B.}
	 \label{fig:colour}
\end{figure*}

The ZTF data does not cover the first brightening event because observations began on 29 August 2018 (MJD$\sim$58359).  The photometry is also not reliable because the binary system is 
 brighter than the saturation limit. However, visual inspection of r-band images provided by the survey shows a clear brightening of the system with time, along with the appearance of a reflection nebula around the September 2020 observations (MJD$\sim$59093). The nebula is still apparent in the latest available images from March 2026 observations (see Fig. \ref{fig:im_ztf}).

To measure the change in the brightness of the nebula with time, we performed aperture photometry for the location of the nebula (blue circle in Fig. \ref{fig:im_ztf}) and of stars in the field around PR Ori. For the photometry, we selected a sub-sample of r-band images from September 2018 to March 2026. Photometry of the nebula was obtained with a 6\arcsec\,aperture using standard IRAF routines. We selected 14 stars with low scatter in the photometry and compared them to the brightness of the nebula (see Fig. \ref{fig:im_ztf}). The photometry is consistent with the nebula becoming brighter between March and September 2020. The brightness rises until approximately October 2023 and has remained at a similar level until the latest available images from ZTF.  These results are robust to the use of different aperture sizes and different locations in the nebulosity. The gap between ZTF observations during the initial rise prevents specific measurements of a time lag \citep[e.g.][observe a $\sim50$ d delay between the outburst and the brightening of the nebula in V1647 Ori]{2004Briceno}.


\subsection{The outburst in mid-infrared photometry}\label{ssec:midir_phot}

The mid-IR photometry shows that the PR Ori system was constant\footnote{Figure \ref{fig:lc} shows the {\it Spitzer}/IRAC magnitude after summing the A and B flux and converting the magnitudes to the WISE passbands, using equations from \citet{2014Antoniucci}.} until the initial burst. 
The {\it NEOWISE} data of the PR Ori system follows the behavior observed in the optical light curve of PR Ori B (see Fig.~\ref{fig:lc}). The first brightening event has $\Delta W1=0.5$, $\Delta W2=0.6$~mag and lasts for roughly $\sim 800$ days, while the second brightening had a much larger amplitude and leveled off with $\Delta W1=2$, $\Delta W2=2.2$~mag  at the end of the NEOWISE mission.

These brightness changes are attributed to variability of PR Ori B. The binary was resolved by {\it Spitzer}/IRAC in the earliest epoch in February 2004, with roughly equal fluxes at 3.5 and 4.5\,$\mu$m \citep{2018Reipurth}.  
Although WISE lacks the angular resolution to resolve the two components, the centroid of the mid-IR emission shifts to PR Ori B as the system gets brighter (see Fig. \ref{fig:mid_ir}), as expected for a brightening of PR Ori B.

Figure \ref{fig:colour} presents a $W1$ versus $W1-W2$ colour-magnitude diagram (CMD) of the combined A+B emission and that of PR Ori B.  After removing the emission from PR Ori A (using the 3.6 and 4.5 $\mu$m fluxes of PR Ori A from \citealt{2018Reipurth}), the amplitude of the outburst to $\Delta W1=3.6$ and $\Delta W2=2.8$~mag and the
$W1-W2$ color gets bluer as the system gets brighter.
The combined emission shows little colour variation, because PR Ori A has a blue $W1-W2$ color and is brighter than the quiescent emission from PR Ori B in $W1$.  

\subsection{The outburst in near-infrared photometry}\label{ssec:nearir_phot}

The limited photometry in the near-IR is consistent with the optical and mid-IR light curves.  PR Ori B  experiences high amplitude variability, while 
PR Ori A remains relatively constant at K$_{\rm s}\sim8.4$~mag.
Comparison of 2MASS photometry obtained in November 2000 with photometry from SpeX observations obtained in February 2023 shows an outburst with amplitudes $\Delta J=1.69$, $\Delta H=1.76$, and $\Delta K_{\rm s}=2.14$~mag for PR Ori B.

\section{Spectroscopic confirmation of an FU Ori Eruption}\label{sec:irspec}

\subsection{The bright viscous disk}
The photometric brightening indicates a large ongoing eruption.
In this section, we establish that this eruption is an FU\,Ori outburst.
In a viscous disk with a high accretion rate, the optical emission will be dominated by the hottest temperatures while the near-IR emission is produced by cooler gas at larger radii in the disk.  This difference leads to the cleanest criteria for classification of FU\,Ori objects, an optical spectrum of an FG supergiant and a near-IR spectrum of an M supergiant star, including strong CO and H$_2$O absorption bands \citep[e.g.][]{miller2011,2018Connelley}. 

Infrared and optical spectra of PR Ori B demonstrates that the emission is now dominated by a viscous disk.  In contrast, the near-IR spectrum of PR Ori B obtained prior to the burst was dominated by the stellar photosphere, while the optical emission of PR Ori B was dominated by scattered light from PR Ori A.

\begin{table}
	\caption{Equivalent Widths of prominent lines}
	\label{tab:app}
\resizebox{\columnwidth}{!}{
   	\begin{tabular}{lccc} 
		\hline
ID & Na I (${\rm \AA}$) & Ca I (${\rm \AA}$) & $^{12}$CO (${\rm \AA}$)\\
\hline
PR Ori A (04-Feb-12)& 2.5$\pm$0.8 & 3.2$\pm$1.1 & 11.5$\pm$4.4\\
PR Ori A (29-Nov-25)& 2.5$\pm$0.2 & 3.2$\pm$0.2 & 12.8$\pm$0.9\\

PR Ori B (21-Jan-13)& 1.8$\pm$0.8 & 2.2$\pm$1.1 & 8.7$\pm$4.5\\

PR Ori B (10-Feb-23)& 0.7$\pm$0.5 & 1.1$\pm$0.2 & 15.6$\pm$0.8\\
PR Ori B (23-Dec-23)& 0.7$\pm$0.3 & 1.2$\pm$0.4 & 19.6$\pm$1.6\\
PR Ori B (20-Aug-25)& 0.6$\pm$0.3 & 1.2$\pm$0.4 & 20.7$\pm$1.5\\
PR Ori B (29-Nov-25)& 0.9$\pm$0.2 & 1.4$\pm$0.3 & 21.3$\pm$1.1\\

V2775 Ori$^{\dagger}$ & 1.6$\pm$0.4 & 1.7$\pm$0.3 & 26.0$\pm$0.7\\
SSTgbs2147 (SE)$^{\dagger}$ & 2.1$\pm$1.0 & 2.0$\pm$0.8 & 22.0$\pm$2.2\\
SPICY97855$^{\ddagger}$ & 1.4$\pm$0.5& 0.2$\pm$0.9 & 14.4$\pm$0.9\\
SPICY99341$^{\ddagger}$ & 1.5$\pm$0.5 & 1.1$\pm$0.9 & 14.8$\pm$0.9\\
SPICY100587$^{\ddagger}$ & 1.8$\pm$0.5 & 1.2$\pm$0.9 & 20.5$\pm$1.1\\
\hline
\multicolumn{4}{l}{$\dagger$ Values taken from \citet{2024Ashraf}}\\
\multicolumn{4}{l}{$\ddagger$ Values taken from \citet{2023Contreras_b}}\\
	\end{tabular}}
\end{table}


\begin{figure*}
    \resizebox{\textwidth}{!}{\includegraphics[angle=0]{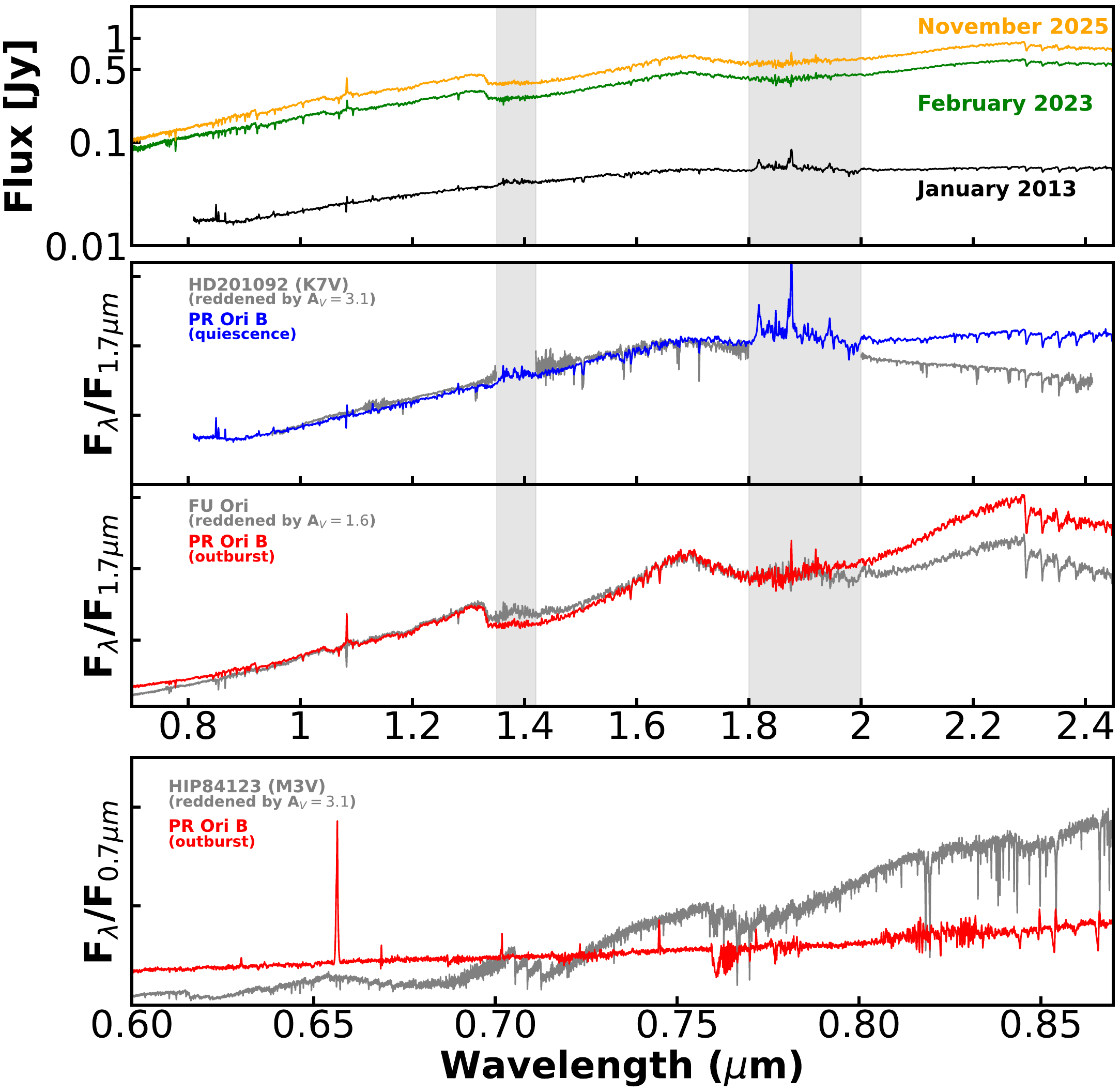}}
	  \caption{(Top) Near-IR spectrum of PR Ori B at different epochs (labeled in the figure). (Middle, top) Near-IR spectrum of PR Ori B during quiescence (January 2013, blue solid line). In the figure we also show the spectrum of HD 201092 (gray solid line), a K7V star, taken from the IRTF spectral library \citep{2009Rayner}. The latter was reddened by A$_{V}=3.1$~mag to match the quiescent spectrum of PR Ori B. (Middle, bottom) Near-IR spectrum of PR Ori B during outburst (November 2025, red solid line). The gray line shows the IRTF/SpeX spectrum of FU\,Ori, a known FUor \citep[][]{2018Connelley}. The spectrum (obtained as part of a separate programme) is reddened by A$_{V}=1.6$~mag to match the outburst spectra of PR Ori B. (Bottom) Optical spectrum of PR Ori B during outburst (red solid line). The figure also shows the optical spectrum of HIP84123, a M3V star, taken from the X-Shooter spectral library \citep{2022Verro}. The latter was reddened by A$_{V}=3.1$~mag for consistency. In the figures, the grey-shaded areas mark regions strongly affected by telluric lines.}
	 \label{fig:spec3}
\end{figure*}

\begin{figure}

\resizebox{\columnwidth}{!}{\includegraphics[angle=0]{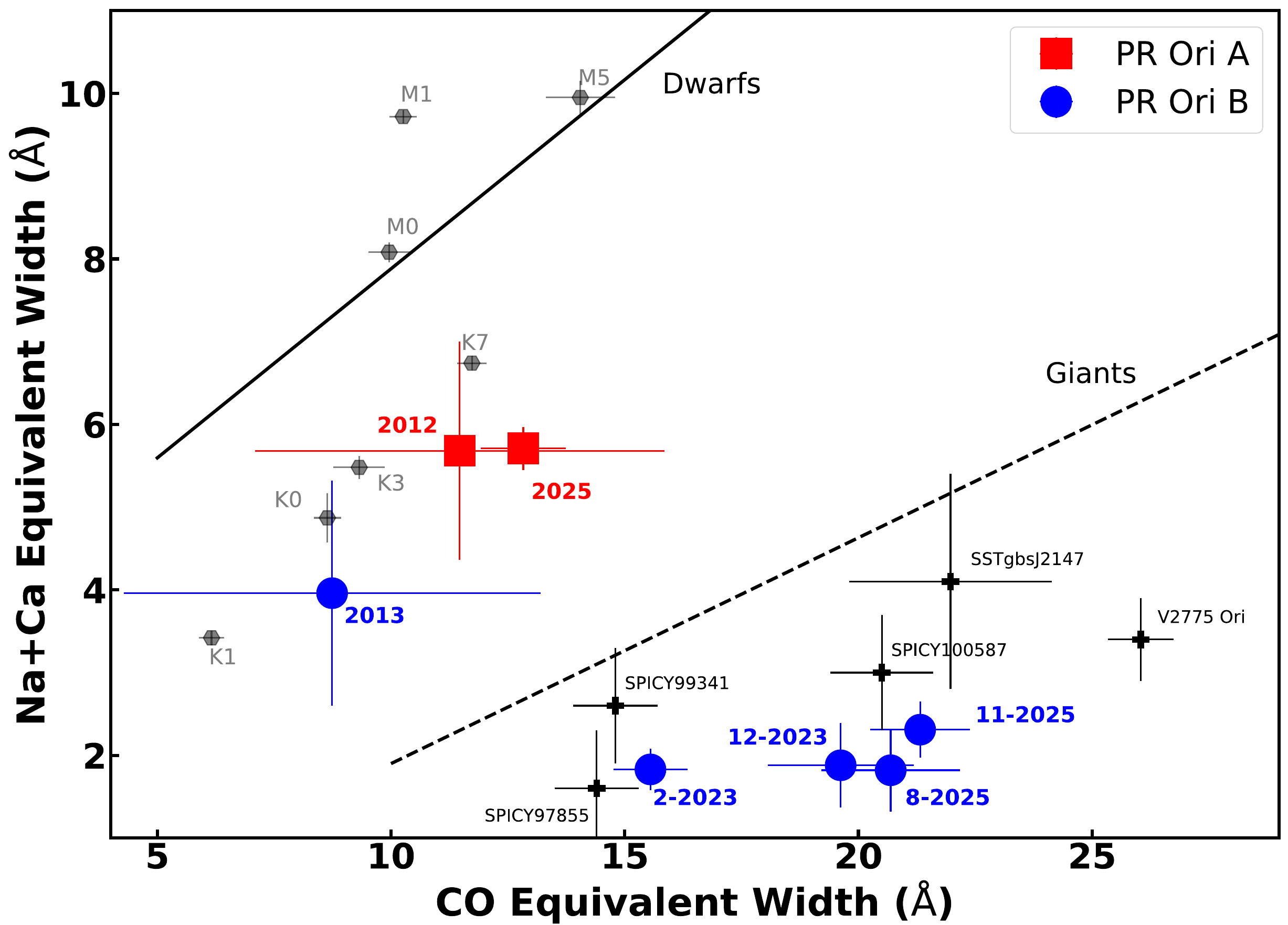}}
	 \caption{Equivalent Width of Na I$+$Ca I versus $^{12}$CO for PR Ori A (red squares) and PR Ori B (blue circles) measured at different dates (marked in the figure). The values for a sample of known FUors (see Table \ref{tab:app}) are also presented in the figure. The expected values for main-sequence and giant stars are shown by black solid and dashed lines, respectively.}
	 \label{fig:ew2}
\end{figure}

The 2023 and 2025 observations of PR Ori B in outburst exhibit strong $^{12}$CO ($\Delta\nu=2$) ro-vibrational absorption bands, along with weak absorption from \ion{Na}{1} (2.21$\mu$m) and \ion{Ca}{1} (2.26$\mu$m). The EWs of \ion{Na}{1}$+$\ion{Ca}{1} versus $^{12}$CO of PR Ori B (see Fig.\ \ref{fig:ew2} and Table~\ref{tab:app}) are located in the region of FUors \citep{2010Connelley}, far from the locus of dwarfs and also discrepant from giant stars. The H-band spectrum shows a triangular shape due to absorption in water vapor bands on both sides of the band (see Fig. \ref{fig:spec3}). These features, which are similar to those of a mid-M type star, are all consistent with the FUor classification criteria established by \citet{2018Connelley}.


The optical spectrum is produced by hotter gas than the near-IR spectrum. Figure \ref{fig:spec3} shows that, at low-resolution, the red spectrum is mostly featureless. The TiO bands at 7140 and 7600 \AA\ are not detected, though they would be prominent for an M-type star, which is in conflict with the apparent spectral type at near-IR wavelengths. The change of spectral type with wavelengths is another defining characteristic of FUor outbursts \citep{2018Connelley}.

\begin{figure*}
\centering
\resizebox{1.9\columnwidth}{!}{\includegraphics[angle=0]{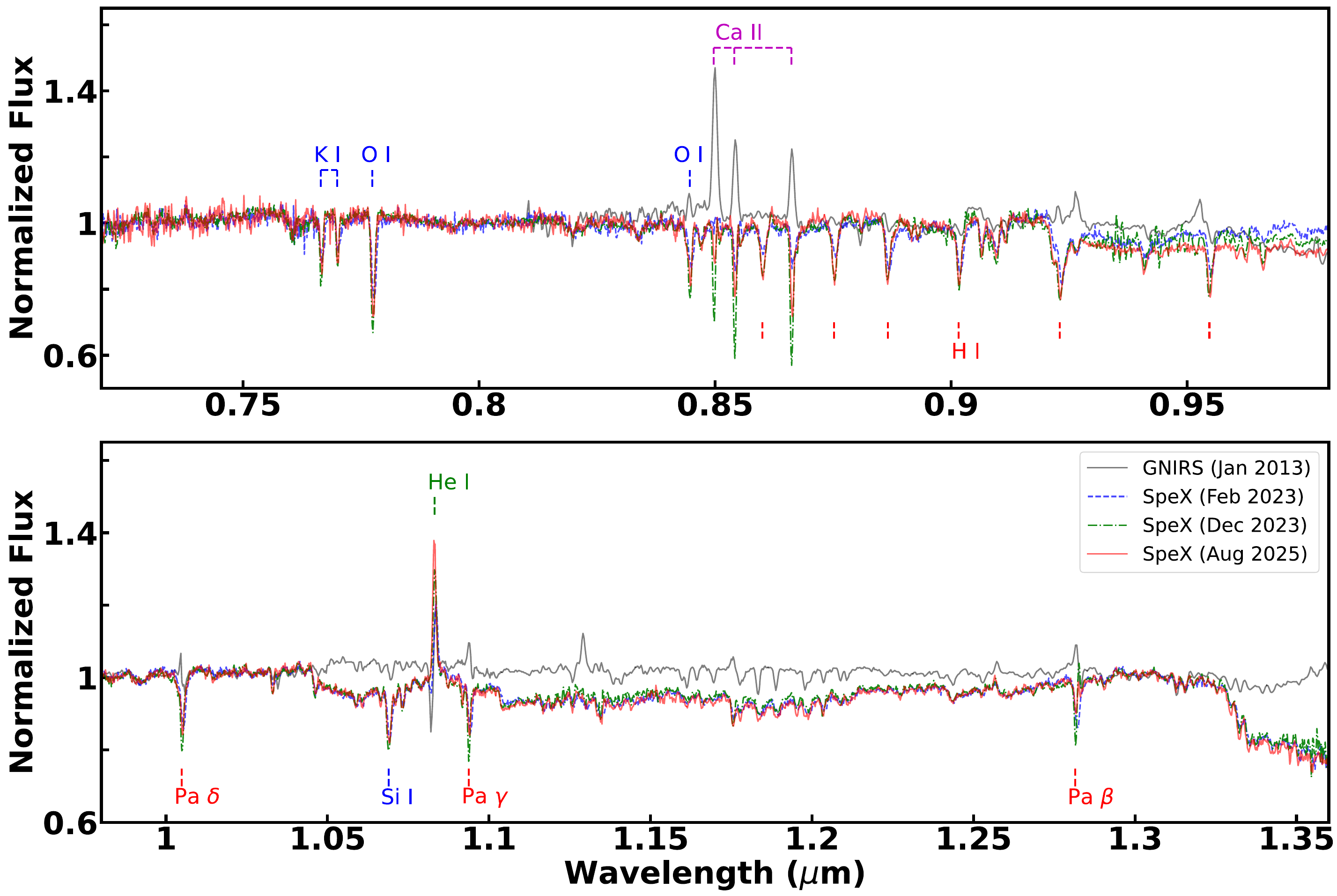}}
	 \caption{Spectrum of PR Ori B between 0.7 and 1.4 $\mu$m. Lines that are typically observed in FUors are marked in the figure.}
	 \label{fig:jspec}
\end{figure*}

During the outburst, H Paschen and Brackett lines in the red-optical, J-band, and K-band spectra are all seen in absorption (Fig.~\ref{fig:jspec}), as expected for FU\,Ori objects \citep[e.g.][]{2018Connelley,2024Guo}.  However, emission in H$\alpha$, and \ion{He}{1}  1.083 $\mu$m lines are discrepant from most previous FU\,Ori objects.  The \ion{He}{1} line in particular is commonly seen in absorption and associated with winds \citep{2018Connelley,2023Ghosh}. Some of this emission may be nebulosity or produced in a different component. The \ion{He}{1} line shows a P\,Cygni profile in the 2013 quiescent spectrum, but is seen purely in emission during outburst. The emission line flux increases by a factor of 45 between quiescence and outburst.

Both the near-IR and optical spectra during the outburst differ from the progenitor spectra described by \citet{2018Reipurth}.  
The pre-outburst near-IR spectrum of PR Ori B is classified as that of a K7$\pm$2 star (see the comparison in Figure \ref{fig:spec3}).  The equivalent width ratios for \ion{Ca}{1}, \ion{Na}{1}, and $^{12}$CO of the January 2013 observations place it in the dwarf sequence and not the FU\,Ori sequence (see Figure \ref{fig:ew}). 
  At shorter wavelengths, the region from 6000-6200 \AA\ showed many photospheric features before the burst, which was the scattered light from PR Ori A  (see Fig.~\ref{fig:hiresspecs}) but only continuum after the burst.

\begin{figure*}
\centering
\resizebox{1.9\columnwidth}{!}{\includegraphics[]{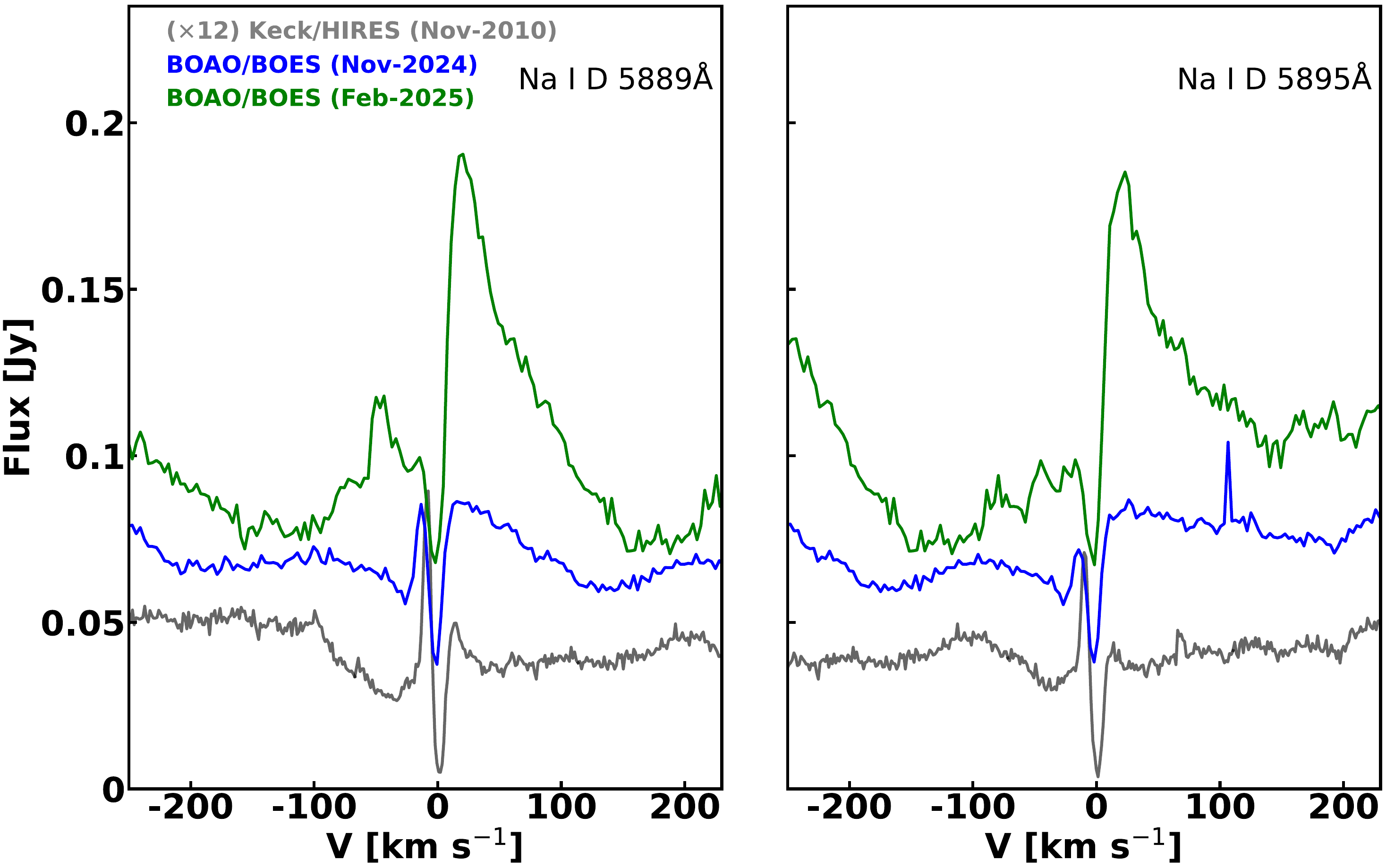}}
\resizebox{1.9\columnwidth}{!}{\includegraphics[]{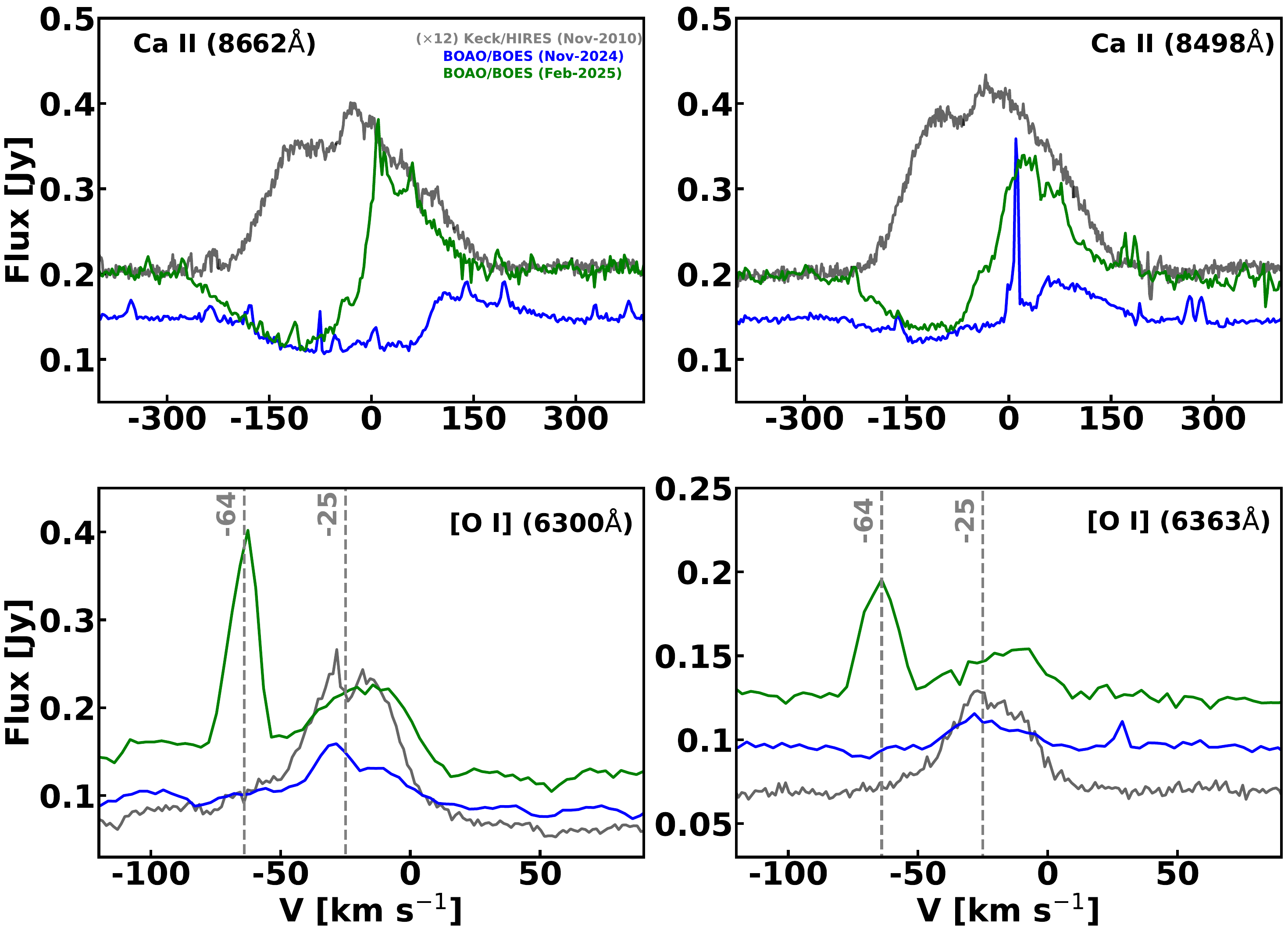}}

	 \caption{Comparison between the high-resolution quiescent spectrum (Keck/HIRES) and the outburst spectra (BOAO/BOES) of PR Ori B for the Na I D ({\it top}), Ca II ({\it middle}) and  [O I] ({\it bottom}) lines.}
	 \label{fig:hiresburst}
\end{figure*}

\subsection{Winds during quiescence and outburst}

During both quiescence and outburst, the PR Ori B spectrum shows emission in a suite of lines, including H$\alpha$, [\ion{O}{1}], the \ion{Ca}{2} IR triplet, and \ion{He}{1} $\lambda10830$. The [\ion{O}{1}] forbidden lines demonstrate the change in wind emission, while the origin of some other lines are unclear.  In this subsection we compare line profiles between quiescence and outburst to assess their origin.

  The line strengths are best measured in the Keck and NGPS observations, which placed both PR Ori A and PR Ori B on the detector and resolved them.  The fluxes are then calibrated with the unchanged spectrum of PR Ori A. The line profiles are best evaluated by comparing the pre-burst HIRES observations with the two high-resolution outburst spectra obtained by BOES (see Fig.\ \ref{fig:hiresburst} for \ion{Na}{1} D lines, the \ion{Ca}{2} IR Triplet, and the forbidden lines [\ion{O}{1}] $\lambda6300,6363$).  However, both BOES 
  high-resolution spectra were obtained on nights with poor seeing ($2\farcs3$-3\arcsec), so emission from both PR Ori A and B is in the spectrum.

From quiescence\footnote{The quiescent emission lines are measured after subtracting off the PR Ori A spectrum from the PR Ori B spectrum (see complications in the PR Ori B quiescent spectrum due to light from source A reflected in the B environment in \S 3).} to outburst, the \ion{Ca}{2} emission increased in flux by a factor of 1.7 and 2.3 (for the 2010 and 2011 observations, respectively), the [\ion{O}{1}] $\lambda$6300,6363 lines increased by a factor of 7, and H$\alpha$ decreased by a factor of 2.2. If the emission is produced in a spatially extended outflow, then the modest difference in \ion{Ca}{2} may be caused by seeing rather than real variability.   The decrease in H$\alpha$ emission may be caused by an underlying absorption component expected for an FU\,Ori object.

The increase in [\ion{O}{1}] emission is substantial enough to point to a real change in line luminosity, either because of an increase in mass loss rate or a stronger radiation field.  The [\ion{O}{1}] lines display a low velocity component, likely related to a disk wind \citep[e.g.][]{2014Natta,banzatti19}, during quiescence and outburst. 
The relative strength of the emission related to the low velocity component decreases during the outburst. The BOES observations from February 2025 also display a high velocity component at $\sim-$64 km s$^{-1}$, 
possibly revealing the presence of a jet or strong outflow \citep{2016Ninan,2022Rab} that was not present in November 2024.

 The \ion{Na}{1} D lines during quiescence show the blueshifted ($-40$ km s$^{-1}$) and redshifted (+200 km~s$^{-1}$) absorption components described by \citet{2018Reipurth}.  After subtracting off the scattered light emission, the blueshifted absorption extends to roughly 100 km~s$^{-1}$.
During the outburst, a P\,Cygni profile develops (as evident in the BOES spectrum from February 2025), with blueshifted absorption extending to $-200$ km s$^{-1}$ and redshifted emission at 
$\sim17$ km s$^{-1}$.

The \ion{Ca}{2} IR triplet lines are covered in both low- (IRTF, NGPS) and high-resolution (Keck/HIRES and BOAO/BOES) observations. 
For accreting young stars, emission in the \ion{Ca}{2} IR triplet is usually associated with magnetospheric accretion \citep{micolta23}, while for FUors they are often seen in absorption and associated with outflows \citep{2015Lee}.
Consistent with these expectations, the lines are in emission during quiescence and have P\,Cygni profiles in the February and October 2025 observations (Fig. \ref{fig:hiresburst}). \ion{Ca}{2} P\,Cygni profiles are also commonly observed in the spectra of FUors \citep{2015Lee,2016Ninan}

The GALAH spectrum of PR Ori B, obtained with a large fiber, shows evidence of outflows, as seen from the detection of \ion{O}{1} absorption and \ion{K}{1} with P Cygni profile at 7700-7800${\rm \AA}$. The spectrum also shows forbidden emission from [\ion{S}{2}] and [\ion{Fe}{2}], which are likely associated with both a jet and the nebulosity.


\section{Properties of PR Ori B before and during the outburst}\label{sec:pr_properties}

In this section we summarize the evidence for PR Ori B as an FU\,Ori object and compare its characteristics to other FU\,Ori objects.   Briefly, PR Ori B is classified as an FU\,Ori object for the following reasons:

\begin{itemize}
    \item The optical brightening of $\sim 5$ mag in {\it Gaia} G band and mid-IR brightening of $\sim3$ mag in W1 and W2 agree with the expectations of an accretion-driven outburst.

\item The near-IR spectra during the outburst show strong absorption in $^{12}$CO and H$_{2}$O bands and \ion{H}{1} lines, typical spectroscopic characteristics of FU\,Ori objects.

\item The optical spectrum is produced by hotter gas than the infrared spectrum, as expected for emission from a viscous disk.

\item The increased luminosity of PR Ori B illuminates a bright nebulosity, one of the historic characteristics of FU\,Ori objects.

\end{itemize}

The only spectroscopic difference between PR Ori B and bona fide FU\,Ori objects is the presence of emission in H$\alpha$ and \ion{He}{1} $\lambda10830$.  However, other H lines are detected in absorption, as expected.  Emission in \ion{He}{1} $\lambda10830$ is also detected in the FU\,Ori object Gaia21bty \citep{2023Siwak} and may be produced in a strong outflow.

In this section, we derive the properties of PR Ori B in an FU Ori outburst and of the progenitor.
The accretion rates for FU\,Ori outbursts are typically measured either from the maximum temperature \citep[e.g.][]{carvalho23} or from SED models \citep[e.g.][]{2023Nagy}, or even more simply just from the absolute magnitude in specific bands \citep{2022Liu,2024Carvalho}.  These approaches first require measurements of the extinction.

\subsection{Estimates of extinction during the outburst}

The extinction A$_{V}$ of PR Ori B is estimated by comparing photometry and flux-calibrated infrared spectra to other FU Ori objects, and separately through disk model fits to the optical-IR SED.  These methods yield discrepant results, which are left unresolved in this paper.   Accretion rate estimates in the next section are provided for this full range of extinctions.

{\it Near-IR photometry:}  During outburst PR Ori B shows $J-K_{\rm s}=1.76\pm0.02$~mag. This is $0.40\pm0.04$ mag redder than the $J-K_{\rm s}=1.36\pm0.03$ color of FU\,Ori from 2MASS photometry. Using the excess of 0.4 mag and the extinction law of \citet{2024Gordon}, we estimate the reddening towards PR Ori B as A$_{V}=(0.4/0.165)+{\rm A}_{V}{\rm (FU~Ori)}$. Taking ${\rm A}_{V}=1.5\pm0.2$~mag for FU\,Ori \citep{2018Connelley}, we determine A$_{V}=3.9\pm0.3$~mag for PR Ori B.

{\it Near-IR spectroscopy:} following the procedure of \cite{2018Connelley}, we redden the spectrum of FU\,Ori \citep[taken from the observations in][]{2018Connelley} until we match that of PR Ori B. A value of ${\rm A}_{V}=1.6\pm0.2$~mag provides an acceptable match. Finally, we add the extinction towards FU\,Ori of ${\rm A}_{V}=1.5\pm0.2$~mag to determine ${\rm A}_{V}=3.1\pm0.3$~mag.

{\it Model SED fits to optical and near-IR photometry:}  Model fits to optical to near-IR fluxes, described below, lead to models with $A_V$ between 1 and 1.7 mag.  In addition, these models require inclinations that are close to $\sim$80\degr, larger than the value derived from the analysis of ALMA data (Section \ref{sec:Obs_datareduction}) and likely inconsistent with the outflow seen only in one lobe (Figures~\ref{fig:image} and~\ref{fig:im_ztf}).  Reproducing the optical photometry would require calculating the radiative transfer, including the scattering surfaces \citep[see for example SED models in][]{whitney03}. Model fits with $A_V=3.1$ mag fail to recover the optical SED but fit the near-IR photometry well, consistent with the other two methods (see discussion in Section \ref{ssec:62}).

\subsection{Estimates of accretion rate during the outburst}\label{ssec:62}

The viscous heating of a disk is described from the theoretical framework of \citet{shakura73}.  The accretion rate can then be measured from the energy released with models or with measurements of the maximum temperature in the inner disk.  In this paper, we use models, either directly or from bolometric corrections based on models.

Table~\ref{tab:accretionrates} shows the accretion rates estimated from individual bands \citep{2024Carvalho}, the WISE bands \citep{2022Liu}, and full fits to optical and infrared photometry (following similar implementations by \citealt{2016Kospal} and \citealt{2023Nagy}).  Most of the accretion luminosities estimates fall between 23--53 $L_\odot$, leading to accretion rates of $\sim 4\times10^{-6}$ to $\sim 10^{-5}$ M$_\odot$ yr$^{-1}$ (with $M_*=0.85$ M$_\odot$, since $L_{\rm acc}\propto M_{\rm acc}M_*$).

The SED model is calculated by integrating black-body emission in concentric annuli between the inner disk radius and the outer disk radius.  The disk uses a fixed inner radius of $2.3$ R$_\odot$, based on the estimate of stellar radius from the quiescent luminosity and effective temperature. We fit the SED using various values of the inclination angle. We find that inclination angles as large as 80$^\circ$ are required to fit the flux of PR Ori B.  This is larger than the 64$^\circ$ derived by \citet{2018Reipurth}, based on the inclination of the HH object, and from our ALMA observations (47$^\circ$, see Section \ref{sec:alma}).

\begin{figure}
    \centering
    \includegraphics[width=0.9\columnwidth]{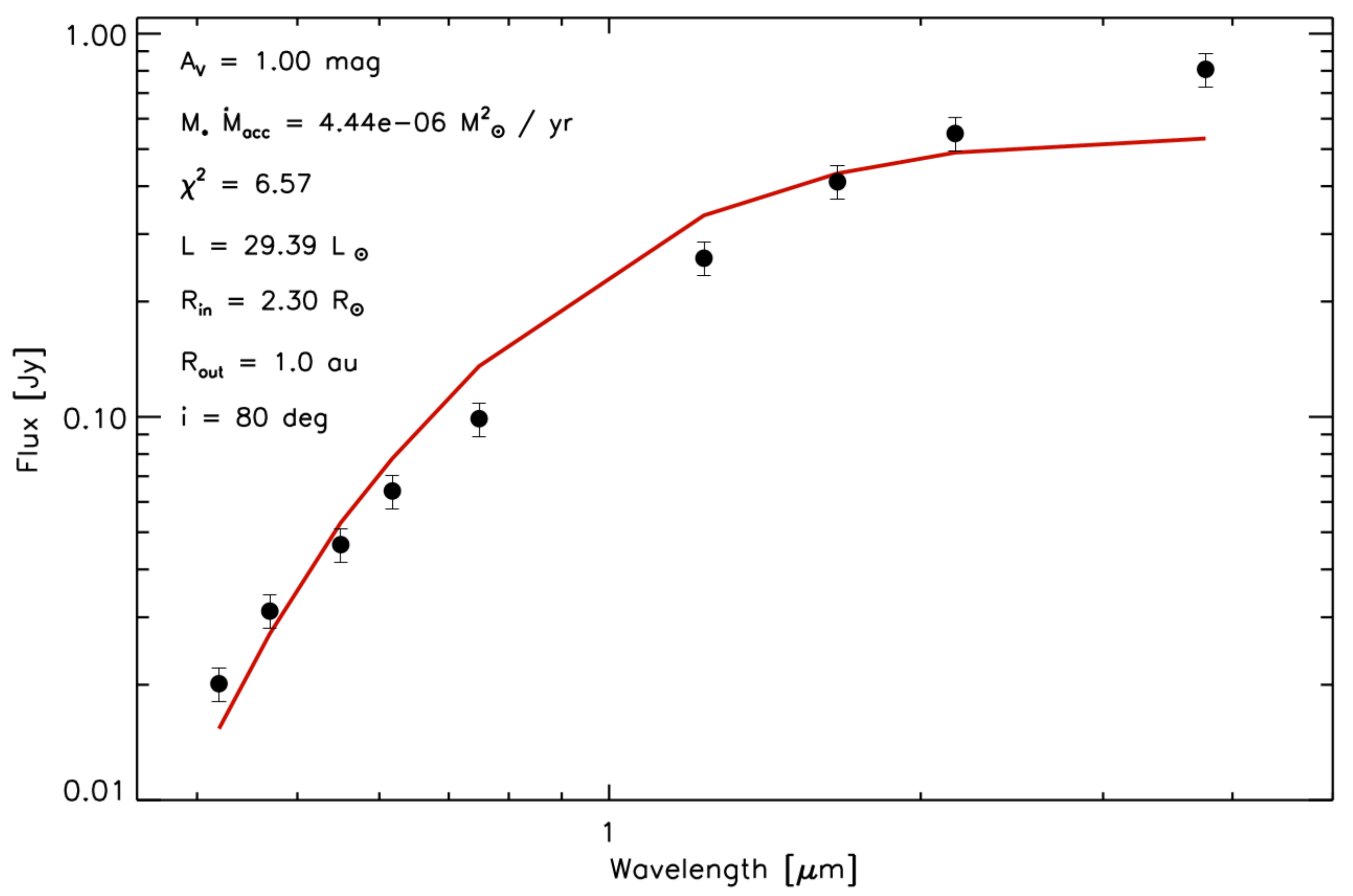}
    \includegraphics[width=0.9\columnwidth]{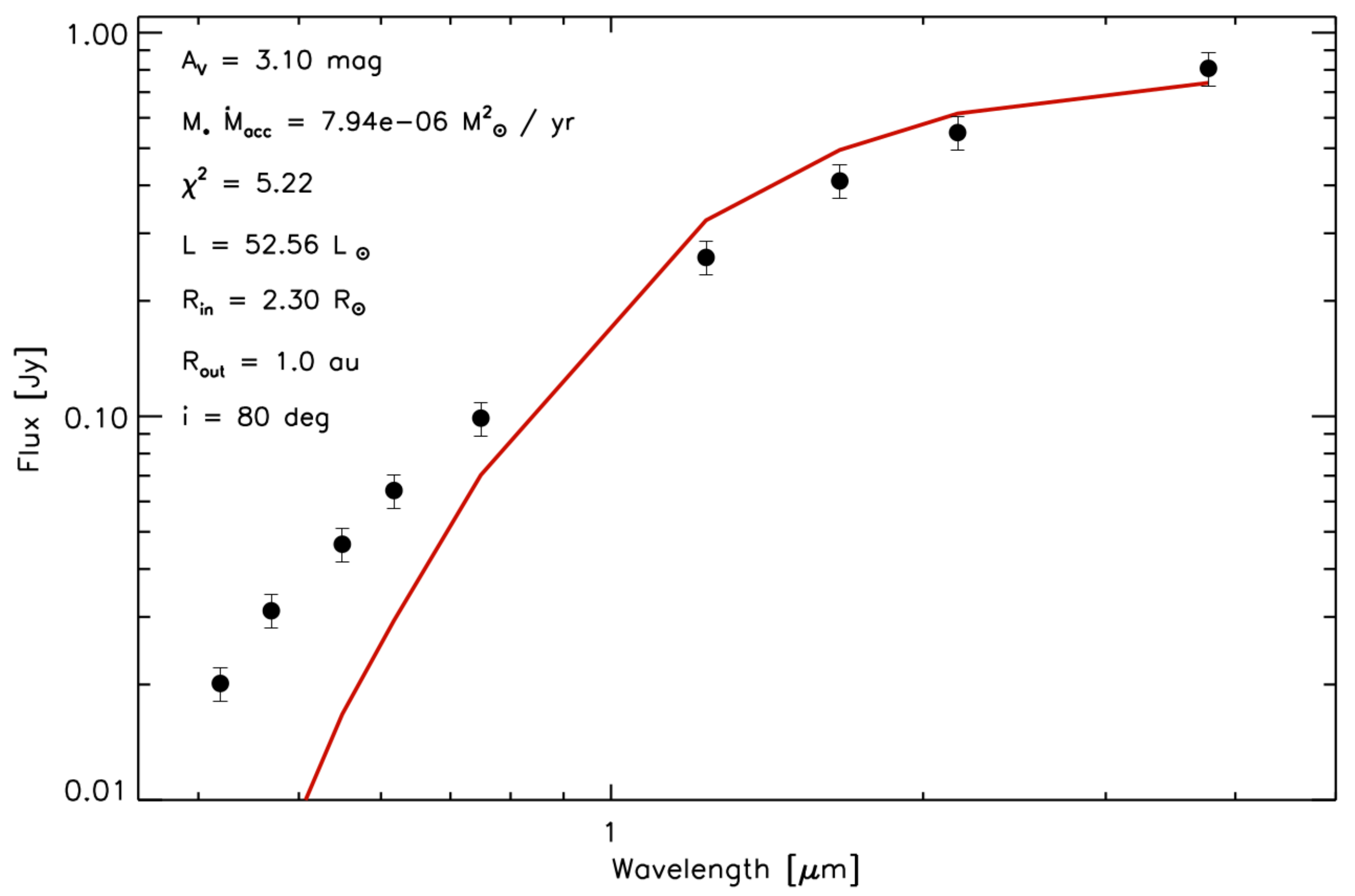}
    \caption{Optical and near-infrared SED of PR Ori B at the epoch of 2023 February 10.
    The solid curve shows the result of the accretion disk model.}
    \label{fig:prorib}
\end{figure}

The SED fits with low $A_V$ reproduce optical emission but overpredict the infrared, including $JHK$ colors that are too blue.  The fits with $A_V=3.1$ mag provide good fits to $JHK$ but severely underpredict optical emission (see Fig. \ref{fig:prorib}).  As a consequence, changing the $A_V$ does not have as significant effect as would be expected in the accretion rate measurements. 

The failure to reproduce both the optical and infrared emission may point to either a deficiency in the disk model or unsteady accretion in the disk. Fits of this model to other FUor outbursts have been much more successful in reproducing the optical-IR SED 
\citep[e.g.][]{2019Kun,2023Nagy}.
If the accretion rate in the innermost disk is higher than at slightly larger radii, the high temperature emission would produce optical emission but less infrared emission.  In addition, most models underpredict the $L$-band emission, perhaps indicating an extra envelope component. The non-negligible contribution of scattered light to optical data \citep[][see e.g. V883 Ori in]{2025Carvalho} could also explain the problems with the fits.

   \begin{table}[!ht]
   \caption{Accretion rate estimates from different methods}
   \label{tab:accretionrates}
  \resizebox{\columnwidth}{!}{\begin{tabular}{lccccccc}
MJD & Type & $A_V$ & $i$ &$L_{\rm acc}$ & $\dot{M}$ & Ref\\
\hline
60362 & W1=6.2 , W2=5.6 & 3.1 & n/a & 23 & $\sim7.4 \times 10^{-6}$ & Liu22 \\
60331 & G=11.3 & 3.1 & 64$^\circ$& 43 & $\sim4.5 \times 10^{-6}$ & CH24 \\
60362 & W1=6. & 3.1& 64$^\circ$ & 26 & $\sim4.6 \times 10^{-6}$ & CH24 \\
60362 & W2=5.3 & 3.1 & 64$^\circ$ & 27 & $\sim4 \times 10^{-6}$ & CH24 \\
59985 & SED & 2.2& 80$^\circ$ & 47 &  $\sim8.1 \times 10^{-6}$ & K16\\
59985 & SED & 3.1& 80$^\circ$ & 53 & $\sim9.1 \times 10^{-6}$ & K16\\
59985 & SED & 1.0 & 80$^\circ$& 29 & $\sim5 \times 10^{-6}$& K16\\
59985  & SED & 1.7& 80$^\circ$ & 79 & $\sim1.4 \times 10^{-5}$& K16\\
\hline
\multicolumn{7}{l}{Liu22: \citet{2022Liu}; CH24: \citet{2024Carvalho};  }\\
\multicolumn{7}{l}{K16: \citet{2016Kospal}}\\
\end{tabular}}
\end{table}

\subsection{The evolution of the outburst}

The rise to maximum light does not resemble the smooth rise observed in the majority of known FUors.
The light curve from PR Ori B includes two fades.  The first occurs from $MJD=58000-58700$ (September 2017 to August 2019), when PR Ori B fades by $\Delta G=3.8$~mag and $\Delta W1=0.8$~mag.  The mid-IR brightness change cannot easily be explained by extinction.  Using the extinction curve of \citet{2024Gordon} with A$_{G}/$A$_{W1}\sim21.1$, the $G$ band fade should lead to a change in $W1$ of 0.18 mag, much less than the 0.8 mag that is observed. Although variable extinction could play a role in this drop in flux, the large amplitude indicates that the first fading is likely driven by a drop in the accretion rate. Similar drops in accretion rate after the initial outburst have been observed in other FUors such as HBC 722 \citep[e.g.][]{2016Kospal}.

PR Ori B starts to become brighter again after MJD$\sim58000$ (September 2017). In the mid-IR, the source getting brighter and bluer, consistent with an increase in the accretion rate of the system \citep[e.g.][]{2022Liu}. Then, PR Ori B shows a second optical fading event at MJD$\sim59400$ (July 2021), with $\Delta G=1.7$~mag, at the same time as a sudden increase in mid-IR emission, with $\Delta W1\sim1.2$~mag. Brighter mid-IR emission and fainter optical emission may be explained by an increased scale height of the inner disk, which can lead to higher optical extinction and larger surface area of warm dust (\citealt{2019Bryan,2021Covey}, see also the scenario for near-IR colors during accretion bursts of EC 53, \citealt{2020Lee}), or perhaps a dusty wind \citep[e.g.][]{petrov19,kadam25}

Though uncommon, two recent FU\,Ori objects had temporary decays.
HBC 722 faded by $\sim1.5$~mag over 5 months, which was explained by a decrease in the accretion rate of the system \citep{2016Kospal}. The recently discovered FUor, L222\_78, shows a decline in its optical light curve while remaining constant at infrared wavelengths, caused by an increase in extinction towards the system by A$_{V}=1$~mag \citep{2024Guo_b}. PR Ori B shows both of these phenomena.

There is no evidence of a lag time between the mid-IR and optical light curves of PR Ori B, as predicted in some models \citep[e.g.][]{masley25}. Such delays have been associated with an inside-out outburst in other FUors \citep[Gaia17bpi, Gaia18dvy, ][]{2018Hillenbrand,2020Szegedi-Elek}. However, the detection of lag times shorter than $\sim$1 year may be difficult due to the cadence of WISE observations.

\subsection{Evidence of silicate crystallization during the outburst}

\begin{figure}
\centering
\resizebox{\columnwidth}{!}{\includegraphics[angle=0]{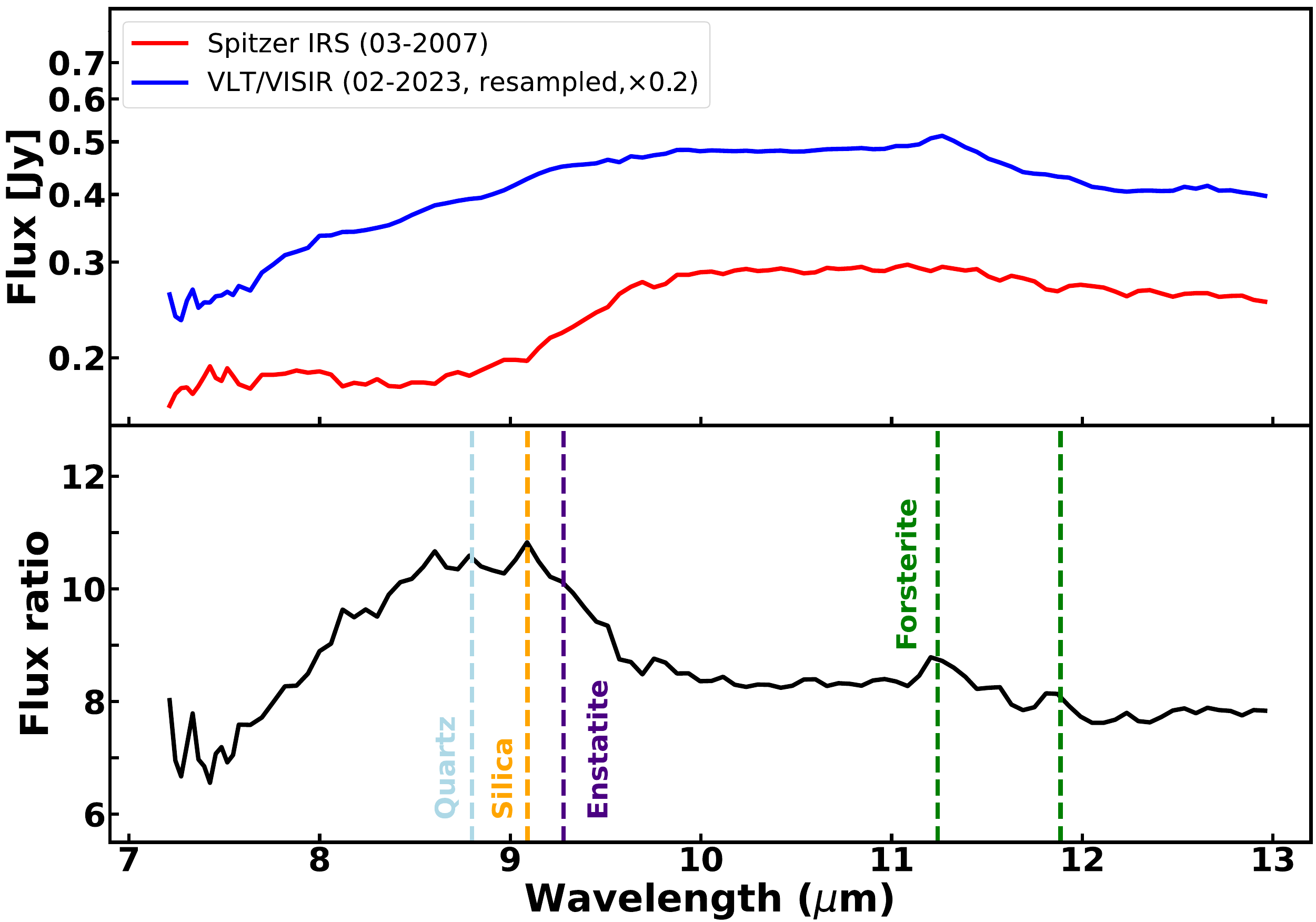}}
	 \caption{(Top) {\it Spitzer}/IRS (red) and VLT/VISIR (blue, scaled by a factor of 0.2) spectra of PR Ori B. The VLT/VISIR spectrum is resampled to the {\it Spitzer}/IRS wavelengths. (Bottom) Flux density ratio between the two observations. The wavelengths of the peaks of crystalline silicates (forsterite, enstatite and quartz), and amorphous silica are marked in the figure.}
	 \label{fig:crystal}
\end{figure}

The presence of crystalline silicates in comets \citep{2006Lisse} suggests that dust processing at high temperatures and subsequent transport towards the outer regions of the disk occurred early in the formation of stars. 
Episodic accretion events may be a pathway for silicate processing in the inner disk. Evidence of crystallization has been detected in the outbursts of EX Lupi \citep{2009Abraham, 2012Juhasz} and EC 53 \citep{2026Lee}. The detection of a nested outflow in EC 53 is consistent with magnetohydrodynamic wind models, providing a mechanism for the outward transport of crystalline silicates \citep{2026Lee}.

We search for evidence of crystallization in the outburst of PR Ori B by comparing the VLT/VISIR observations taken during outburst, with the {\it Spitzer/IRS} quiescent observations from 2007 (Fig.~\ref{fig:crystal}). The VISIR spectrum is resampled to the lower resolution of the {\it Spitzer/IRS} spectrum.

The ratio of VISIR to Spitzer/IRS flux (bottom plot in Fig. \ref{fig:crystal}) shows clear peaks at $\sim$11.2 and $\sim$11.9\,$\mu$m, corresponding to emission by crystalline forsterite (the Mg-rich form of olivine). The enstatite feature at $\sim$9.2\,$\mu$m is not definitively detected.
 The broad emission with a peak at $\sim9\,\mu$m is likely associated with amorphous silica \citep{2009Francis} and crystalline quartz \citep{2013Zeidler}. These changes indicates the increase in emission from silica as it is heated during the outburst and the production of quartz in the hot inner disk.
 
We do not have data at longer wavelengths to discard whether the emission originated from pre-existing crystalline silicates \citep[observations at $\lambda>16\,\mu$m are used as evidence in][]{2009Abraham,2026Lee}. Future JWST observations could help to confirm that we are observing in-situ formation of crystals in the outburst of PR Ori B.

\section{Conclusions}\label{sec:pr_importance}

In this work we have shown that PR Ori B is the driver of the high-amplitude variability that was detected by the {\it Gaia}, ASAS-SN and WISE/NEOWISE surveys towards the binary system, PR Ori. The amplitude of the outburst is $\Delta G=5\,$mag, with a peak accretion rate of $\sim4-14\times10^{-6}$~M$_{\odot}$~yr$^{-1}$.  The luminosity of the star increased by a factor of 10-20. The peak accretion luminosity and rate is lower than average for FU Ori objects \citep[see compilation by][]{2025Contreras_b}.

The striking similarities between the pre-outburst absorption spectra of PR Ori A and B means that we are unable to fully determine the photospheric properties of PR Ori B from existing optical spectra. Although there are complications in the interpretation of the spectra arising from this binary system \citep[see also][]{2026Lee_E}, the near-IR spectra of PR Ori B clearly shows the change from a late K-type star during quiescence to a viscously heated disk during outburst.

FU\,Ori events are rare in Class II objects, with timescales of one per $10^5$\,yr \citep{2019Contreras}.  Similar events in younger Class I and Class 0 protostars may be more common \citep[e.g.][]{scholz13,2019Fischer,2021Park,2022Zakri,2024Contreras} but are optically faint and difficult to confirm and study spectroscopically. Only a few confirmed FU\,Ori objects are located within 1 kpc  (see the catalogs in \citealt{2014Audard}, \citealt{2025Contreras_b} and recent discoveries, for example results from the VVV Survey in \citealt{2024Guo}). PR Ori B is located in Orion, one of the closest and best-studied star-forming regions. \citet{2018Reipurth} found that shocks in the outflow occur at intervals of hundreds of years, which may indicate that PR Ori B has had frequent, repeated bursts in the past.

The discovery of a likely FU\,Ori-type outburst in Orion means that we have extensive archival data of the pre-outburst star and photometry before the start of the burst.  The nearby distance will allow high angular resolution observations to probe the physics in the inner disk.  The distance and low extinction also means that the source is optically bright -- the second brightest FU\,Ori or FU\,Ori-like object known at optical wavelengths \citep[from comparing with known sources in OYCAT,][]{2025Contreras_b}.  PR Ori B is only about one mag fainter than FU\,Ori in $g$-band with ASAS-SN and in $G$ with {\it Gaia}, with the possibility that the eruption brightens in the future.

Future observations will also be powerful in evaluating changes in the disk that result from the increased luminosity. Sub-mm ALMA observations obtained on 2019 Dec 12 (MJD 58829), just months before the sharp rise in the mid-IR, establishes a baseline to evaluate any increase in the dust temperature.  


{







\section{acknowledgments}
This research has made use of the NASA/IPAC Infrared Science Archive, which is funded by the National Aeronautics and Space Administration and operated by the California Institute of Technology. We acknowledge with thanks the variable star observations from the AAVSO International Database contributed by observers worldwide and used in this research. This work has made use of data from the European Space Agency (ESA) mission {\it Gaia} (\url{https://www.cosmos.esa.int/gaia}), processed by the {\it Gaia} Data Processing and Analysis Consortium (DPAC, \url{https://www.cosmos.esa.int/web/gaia/dpac/consortium}). Funding for the DPAC has been provided by national institutions, in particular the institutions participating in the {\it Gaia} Multilateral Agreement. This work is partly based on data obtained with the Infrared Telescope Facility, which is operated by the University of Hawaii under contract 80HQTR24DA010 with the National Aeronautics and Space Administration. This paper makes use of the following ALMA data: ADS/JAO.ALMA2023.1.00561.S ALMA is a partnership of ESO (representing its member states), NSF (USA) and NINS (Japan), together with NRC (Canada), NSTC and ASIAA (Taiwan), and KASI (Republic of Korea), in cooperation with the Republic of Chile. The Joint ALMA Observatory is operated by ESO, AUI/NRAO and NAOJ

CCP was supported by the National Research Foundation of Korea (NRF) grant funded by the Korean government (MEST) (No. 2019R1A6A1A10073437). JEL, CHK, and CCP were supported by the NRF grant funded by the Korean government (MSIT) (grant numbers RS-2024-00416859 and RS-2026-25490557). G.J.H. is supported by general grant 12573031 from the National Natural Science Foundation of China and by National Key R\&D program 2022YFA1603102 from the Ministry of Science and Technology (MOST) of China. 

D.J.\ is supported by NRC Canada and by an NSERC Discovery Grant. A.K. and F.L. were supported by ADVANCED 149943 and 2024-1.2.8-T\'ET-IPARI-CN-2025-00036 grants, which have been implemented with the support provided by the Ministry of Culture and Innovation of Hungary from the National Research, Development and Innovation Fund, financed under the NKKP ADVANCED and 2024-1.2.8-T\'ET-IPARI-CN funding schemes. P.A. and F.L. received funding from the Hungarian NKFIH NKKP project No. K-147380. The operation of the RC80 telescope at Konkoly Observatory has been supported by the GINOP 2.3.2-15-2016-00033 grant of the National Research, Development and Innovation Office (NKFIH) funded by the European Union.
M. Sz. and Zs. N. acknowledge support from the ESA PRODEX contract nr. 4000132054.
Zs. N. acknowledges the Hungarian National Research, Development and Innovation Office grant OTKA FK 146023.
Zs. N. was supported by the J\'anos Bolyai Research Scholarship of the Hungarian Academy of Sciences.
Zs. H., K. L. and N. O. Sz. thank the financial support provided by the undergraduate research assistant program of Konkoly Observatory.
Zs. H., Cs. K. and K. L. were supported by the ``SeismoLab'' KKP-137523 grant of the Hungarian Research, Development and Innovation Office (NKFIH). B.S. was supported by the Hungarian National Research, Development and Innovation Office \'Elvonal grant KKP-143986.
F. C. S. M. received financial support from the European Research Council (ERC) under the European Union's Horizon 2020 research and innovation programme (ERC Starting Grant ``Chemtrip'', grant agreement No 949278).
 N. O. Sz. is supported by the EK\H{O}P-25 university research scholarship program of the Ministry for Culture and Innovation from the source of the National Research, Development and Innovation Fund.
This project has received funding from the European Research Council (ERC) via the ERC Synergy Grant ECOGAL (grant 855130). Views and opinions expressed are however those of the author(s) only and do not necessarily reflect those of the European Union or the European Research Council Executive Agency. Neither the European Union nor the granting authority can be held responsible for them.


\vspace{5mm}
\facilities{IRTF:Spex, WISE}

\bibliography{prori_v3}{}
\bibliographystyle{aasjournal}
}
\end{CJK}

\end{document}